\documentclass[]{aastex631}
\usepackage{ulem}

\newcommand\subs[1]{\textsubscript{#1}}
\newcommand\sups[1]{\textsuperscript{#1}}
\newcommand\rh[1]{\textcolor{black}{{\textit{r}\subs{H}}#1}}

\newcommand\Ju[1]{\textcolor{black}{{\textit{J}}#1}}
\newcommand\ie[1]{\textcolor{black}{{\textit{i.e.,}}#1}}
\newcommand\eg[1]{\textcolor{black}{{\textit{e.g.,}}#1}}

\newcommand\vexp[1]{\textcolor{black}{{\textit{v}\subs{exp}}#1}}
\newcommand\kms[1]{\textcolor{black}{{km\,s$^{-1}$}#1}}

\definecolor{gold}{rgb}{0.64,0.54,0.29}

\received{May 2023}
\accepted{September 2023}
\submitjournal{PSJ}

\shorttitle{ALMA Observations of the DART Impact}
\shortauthors{Roth et al.}

\begin{document}


\title{ALMA Observations of the DART Impact: Characterizing the Ejecta at Sub-Millimeter Wavelengths}

\correspondingauthor{Nathan X. Roth}
\email{nathaniel.x.roth@nasa.gov}

\author[0000-0002-6006-9574]{Nathan X. Roth}
\affiliation{Solar System Exploration Division, Astrochemistry Laboratory Code 691, NASA Goddard Space Flight Center, 8800 Greenbelt Rd., Greenbelt, MD 20771, USA}
\affiliation{Department of Physics, The Catholic University of America, 620 Michigan Ave., N.E. Washington, DC 20064, USA}

\author[0000-0001-7694-4129]{Stefanie N. Milam}
\affiliation{Solar System Exploration Division, Astrochemistry Laboratory Code 691, NASA Goddard Space Flight Center, 8800 Greenbelt Rd., Greenbelt, MD 20771, USA}

\author[0000-0001-9479-9287]{Anthony J. Remijan}
\affiliation{National Radio Astronomy Observatory, 520 Edgemont Rd, Charlottesville, VA 22903, USA}

\author[0000-0001-8233-2436]{Martin A. Cordiner}
\affiliation{Solar System Exploration Division, Astrochemistry Laboratory Code 691, NASA Goddard Space Flight Center, 8800 Greenbelt Rd., Greenbelt, MD 20771, USA}
\affiliation{Department of Physics, The Catholic University of America, 620 Michigan Ave., N.E. Washington, DC 20064, USA}

\author[0000-0003-2474-7523]{Michael W. Busch}
\affiliation{SETI Institute, 189 Bernardo Avenue, Suite 200, Mountain View, CA 94043, USA}

\author[0000-0003-3091-5757]{Cristina A. Thomas}
\affiliation{Northern Arizona University, Department of Astronomy and Planetary Science, P.O. Box 6010, Flagstaff, AZ 86011, USA}

\author[0000-0002-9939-9976]{Andrew S. Rivkin}
\affiliation{Johns Hopkins University Applied Physics Laboratory, 11100 Johns Hopkins Rd., Laurel, MD, 20723 USA}

\author[0000-0002-9820-1032]{Arielle Moullet}
\affiliation{National Radio Astronomy Observatory, 520 Edgemont Rd, Charlottesville, VA 22903, USA}

\author[0000-0003-1450-5299]{Ted L. Roush}
\affiliation{NASA Ames Research Center (Retired), Planetary Systems Branch, MS 245-3, Moffett Field, CA 94035, USA}

\author[0000-0002-8505-4934]{Mark A. Siebert}
\affiliation{Department of Astronomy, University of Virginia, Charlottesville, VA 22904, USA}

\author[0000-0003-3841-9977]{Jian-Yang Li}
\affiliation{Planetary Science Institute, 1700 East Fort Lowell, Suite 106, Tucson, AZ 85719, USA}

\author{Eugene G. Fahnestock}
\affiliation{Jet Propulsion Laboratory, California Institute of Technology, Pasadena, CA 91109, USA}

\author[0000-0001-8417-702X]{Josep M. Trigo-Rodr\'{i}guez}
\affiliation{Institute of Space Sciences (CSIC-IEEC), Campus UAB, Carrer de Can
Magrans s/n, 08193 Cerdanyola del Vallés (Barcelona), Catalonia, Spain}

\author{Cyrielle Opitom}
\affiliation{ Institute for Astronomy, University of Edinburgh, Royal Observatory, Edinburgh EH9 3HJ, UK}

\author[0000-0002-1821-5689]{Masatoshi Hirabayashi}
\affiliation{Department of Aerospace Engineering, Department of Geosciences, Auburn University, Auburn, AL, USA}
\affiliation{Daniel Guggenheim School of Aerospace Engineering, Georgia Institute of Technology, Atlanta, GA, USA}




\begin{abstract}
We report observations of the Didymos-Dimorphos binary asteroid system using the Atacama Large Millimeter/Submillimeter Array (ALMA) and the Atacama Compact Array (ACA) in support of the Double Asteroid Redirection Test (DART) mission. Our observations on UT 2022 September 15 provided a pre-impact baseline and the first measure of Didymos-Dimorphos' spectral emissivity at $\lambda=0.87$ mm, which was consistent with the handful of siliceous and carbonaceous asteroids measured at millimeter wavelengths. Our post-impact observations were conducted using four consecutive executions each of ALMA and the ACA spanning from T$+$3.52 to T$+$8.60 hours post-impact, sampling thermal emission from the asteroids and the impact ejecta. We scaled our pre-impact baseline measurement and subtracted it from the post-impact observations to isolate the flux density of mm-sized grains in the ejecta. Ejecta dust masses were calculated for a range of materials that may be representative of Dimorphos' S-type asteroid material. The average ejecta mass over our observations is consistent with 1.3--6.4$\times10^7$ kg, with the lower and higher values calculated for amorphous silicates and for crystalline silicates, respectively. Owing to the likely crystalline nature of S-type asteroid material, the higher value is favored. These ejecta masses represent 0.3--1.5\% of Dimorphos' total mass and are in agreement with lower limits on the ejecta mass based on measurements at optical wavelengths. Our results provide the most sensitive measure of mm-sized material in the ejecta and demonstrate the power of ALMA for providing supporting observations to spaceflight missions.

\end{abstract}


\keywords{Asteroids (72) --- Near Earth Objects (1092) --- Impact Phenomena (779) --- Radio astronomy (1338) --- Radio interferometry(1346) --- High resolution spectroscopy (2096)}


\section{Introduction}

The Double Asteroid Redirection Test (DART) mission provided the first demonstration of the kinetic impactor planetary defense technique \citep{Daly2023}. Targeting Didymos-Dimorphos, a binary near-Earth asteroid (NEA) system with a heliocentric orbital period $P$ = 2.11 years, the spacecraft impacted the smaller asteroid, Dimorphos, at UT 23:14 on 2022 September 26. The small geocentric distance of the Didymos-Dimorphos system ($\Delta_{min}$ = 0.075 au) at impact and in the days beforehand afforded an opportunity to characterize both the system and the impact ejecta at high sensitivity using ground- and space-based observatories. The impact produced a spectacular plume of ejecta which continuously evolved, eventually developing into a comet-like tail \citep[\eg{}][]{Thomas2023,Li2023,Opitom2023,Graykowski2023}. 

Here we report pre- and post-impact observations of the Didymos-Dimorphos system using the Atacama Large Millimeter/Submillimeter Array (ALMA) with Director's Discretionary Time. We conducted pre-impact observations in order to characterize Didymos-Dimorphos' spectral emissivity at millimeter wavelengths for the first time, and post-impact observations to measure the gas and dust produced in the ejecta plume. Using Didymos-Dimorphos' spectral emissivity at the $\lambda$ = 0.87 mm wavelength of our observations, we were able to calculate the expected flux density of the asteroids on the impact date in order to isolate the ejecta in our post-impact observations. Based on the ejecta particle size distribution measured by the Hubble Space Telescope \citep[HST;][]{Li2023}, the total ejecta mass was dominated by large ($>$ 100 $\mu$m) particles, making our ALMA measurements at $\lambda$ = 0.87 mm the most sensitive observations to the total ejecta mass. In Section~\ref{sec:obs} we detail our observations and data reduction. In Section~\ref{sec:results} we provide our results. In Section~\ref{sec:modeling} we explain our modeling formalism and interpret our analysis in the context of planetary defense and the asteroid population.

\section{Observations and Data Reduction}\label{sec:obs}
\subsection{Pre-Impact 12m Observations}\label{subsec:pre-impact}
On UT 2022 September 15 we targeted Didymos-Dimorphos with one execution of the ALMA 12 m array using the Band 7 receiver in Time Division Mode (TDM) for pre-impact continuum observations centered at 343.5 GHz ($\lambda$ = 0.87 mm) with a total bandwidth of 8 GHz in two polarizations. The observing log is shown in Table~\ref{tab:obslog}. We tracked the asteroids' position using JPL Horizons ephemerides (\#187). Quasar observations were used for bandpass and phase calibration, and J2258-2758 was used to calibrate the flux scale. Weather conditions were excellent (precipitable water vapor, PWV, at zenith 0.32 mm). The spatial scale (the range in major and minor FWHM of the synthesized beam) was 0$\farcs$38 -- 0$\farcs$56 and the channel spacing was 31.25 MHz for continuum windows, leading to a spectral resolution of 27 km s$^{-1}$. The data were flagged and calibrated using standard routines in Common Astronomy Software Applications (CASA) package version 6.5.2 \citep{CASA2022}. Self-calibration of the antenna phases and amplitudes was applied to improve the image signal-to-noise ratio (S/N; Appendix~\ref{sec:selfcal}).

After self-calibration, we generated the final images using the multi-term multi-frequency synthesis (MTMFS) algorithm with Briggs visibility weighting (with a robust value of 0.5) and auto-masking using the standard pipeline threshold parameters specified for continuum measurements with a compact 12 m array \citep{Kepley2020}, followed by primary beam correction. Thermal continuum emission was clearly detected. We transformed the images from astrometric coordinates to projected distances at the asteroids, with the location of the peak continuum flux chosen as the origin. This peak continuum position was in agreement with the predicted ephemeris position of the asteroids to within one synthesized ALMA beam.

\subsection{Post Impact 12m and ACA Observations}\label{subsec:post-obs}
On UT 2022 September 27 we targeted Didymos-Dimorphos in four consecutive (46 minutes each), simultaneous executions of the ALMA 12 m array and the Atacama Compact Array (ACA) using the Band 7 receiver in Frequency Division Mode (FDM) covering frequencies from 344.25 GHz to 347.45 GHz ($\lambda$ = 0.86 -- 0.87 mm) in six non-contiguous spectral windows ranging from 117 MHz to 2000 MHz wide and a total bandwidth of 3 GHz in two polarizations. This correlator setup was chosen to be sensitive to spectral line emission from potential gas-phase ejecta as well as to thermal continuum emission. Based on the presence of silicates (olivenes, pyroxenes) and sulfides (troilite) in S-type asteroids \citep{Lawrence2007}, we targeted the SiO (\Ju{}=8--7), SiS (\Ju{}=19--18), and SO ($J_K$=$8_8$--$7_7$) transitions to test for gas-phase silicon- or sulfur-bearing species in the ejecta. Transitions of two alkali species, KCl ($J$=45--44) and NaCN ($J_{Ka,Kc}$=$22_{3,20}$--$21_{3,19}$), were also covered by our correlator setup.

The channel spacing was 15.625 MHz for continuum windows and 61 kHz -- 122 kHz for spectral line windows, resulting in a spectral resolution of 0.05 -- 0.11 \kms{}. The spatial scale was 0$\farcs$48 -- 1$\farcs$11 for the 12 m array and $2\farcs79$ -- $6\farcs95$ for the ACA. Tracking and calibration were performed as in our pre-impact epoch, including the application of self-calibration to continuum measurements. Weather conditions were excellent (PWV at zenith 0.46--0.47 mm).

Quasar observations were used for bandpass and phase calibration. The following quasars were used to calibrate the flux scale: J0006-0623 for 12 m Executions 1 and 2, J0238+1636 for 12 m Execution 3, J0519-4546 for 12 m Execution 4, J2253+1608 for ACA Executions 1 and 2, J2258-2758 for ACA Execution 3, and J0423-0120 for ACA Execution 4. We examined the ALMA Calibrator Source Catalog for each quasar to assess the absolute flux calibration scale assigned by the pipeline.

All quasars followed smooth trends in their measured Band 7 fluxes between 2022 June and 2022 December, with the exception of J0238+1636 (12 m Execution 3), which showed a steady increase in flux between June (0.54 Jy) and September (1.11 Jy), followed by a downward trend beginning in October (0.9 Jy) through December (0.46 Jy). The automated ALMA pipeline calibration scheme extrapolated along the upward trend to assign a flux scale of 1.22 Jy to J0238+1636 during our 12 m Execution 3, resulting in artificially inflated flux for our Didymos-Dimorphos measurements. We instead applied the value of 1.11 Jy (measured for J0238+1636 on 2022 September 23) to our 12 m Execution 3 observations to better reflect the observed behavior of the flux calibrator. Owing to this calibrator behavior, we conservatively adopted the 10\% uncertainty in absolute flux recommended by the ALMA observatory for our measurements.

Spectral line data were imaged before any self-calibration was applied using the H\"{o}gbom algorithm with natural weighting and auto-masking using the standard pipeline threshold parameters specified for spectral line measurements with a compact 12 m array or the ACA \citep{Kepley2020}, followed by primary beam correction. Natural weighting was chosen to maximize sensitivity to potential faint, extended spectral line emission. Spectral line emission was not detected in individual or summed executions for the 12 m array or the ACA.

We then averaged the spectral line channels to a channel width of 250 MHz to create a channel-averaged continuum Measurement Set for both the 12 m and the ACA data and applied self-calibration as for the pre-impact data. After self-calibration, we generated the final images using the multi-term multi-frequency synthesis (MTMFS) algorithm with Briggs visibility weighting and auto-masking using the standard pipeline threshold parameters specified for continuum measurements with a compact 12 m array or the ACA \citep{Kepley2020}, followed by primary beam correction. Briggs visibility weighting (with a robust value of 0.5) was chosen to balance angular resolution and sensitivity to continuum emission. Thermal continuum emission was clearly detected for all executions of the 12 m array and the ACA. We transformed the images from astrometric coordinates to projected distances at the asteroids, with the location of the peak continuum flux chosen as the origin. This peak continuum position was in agreement with the predicted ephemeris position of the asteroids to within one synthesized ALMA beam.

\subsection{Observing Geometry}\label{subsec:obs-geometry}
Here we detail aspects of the observing geometry and properties of the Didymos-Dimorphos system relevant for our analysis. \cite{Daly2023} calculated volume equivalent sphere diameters of 761 m and 151 m for Didymos and Dimorphos, respectively, based on imaging by the DRACO instrument on DART. Given their sizes and separation distance of $\sim$1.2 km \citep{Naidu2020}, neither asteroid was spatially resolved during our pre-impact or post-impact observations (see Table~\ref{tab:obslog} for the angular resolution).

No mutual events occurred during our pre-impact observations \citep{Scheirich2022}; thus, we can separate each asteroid's contributions to the measured flux density. Additionally, the 12 m array was slightly more extended during our pre-impact epoch and the asteroids were observed at elevation $\sim60\degr$, resulting in a more compact synthesized beam than our post-impact epochs. For our post-impact epochs, a primary eclipse occurred from UT September 26 23:49 -- September 27 00:48, followed by a secondary eclipse from UT September 27 05:46 -- 07:01. The next eclipse did not begin until UT September 27 11:43. Thus, Execution 4 of the 12 m array and Execution 3 of the ACA were conducted during a secondary eclipse, while the remaining executions were outside of any mutual events (Table~\ref{tab:obslog}).

\begin{deluxetable*}{cccccccccccc}[h]
\tablenum{1}
\tablecaption{Observing Log\label{tab:obslog}}
\tablewidth{0pt}
\tablehead{
\colhead{UT Time} & \colhead{Time Post-Impact} & \colhead{\textit{T}\subs{int}} &
\colhead{\textit{r}\subs{H}} & \colhead{$\Delta$} & \colhead{$\alpha$} &  
\colhead{\textit{N}\subs{ants}}  & \colhead{Baselines} & \colhead{PWV} & \colhead{El.} & \colhead{$\theta$\subs{AR}} & \colhead{$\theta$\subs{AR}}  \\
\colhead{(2022)} & \colhead{(hr)} & \colhead{(min)} & \colhead{(au)} & 
\colhead{(au)} & \colhead{($\degr$)} & \colhead{} & \colhead{(m)}  & \colhead{(mm)} & \colhead{($\degr$)} & \colhead{($\arcsec$)} & \colhead{(km)}
}
\startdata
\multicolumn{12}{c}{12m Pre-Impact Observations, 2022 September 15, $\nu$ = 343.5 GHz, $\theta$\subs{MRS} = $3\farcs3$ (236 km)} \\
03:49--03:55 & $\cdot\cdot\cdot$ & 5 & 1.082 & 0.099 & 37.5 & 44 & 15--782 & 0.32 & 56 & 0.56$\times$0.38 & 40$\times$27 \\
\hline
\multicolumn{12}{c}{DART Impact: UT 23:14 2022 September 26} \\
\hline
\multicolumn{12}{c}{Post-Impact 12 m Observations, 2022 September 27, $\nu$ = 345 GHz, $\theta$\subs{MRS} = $4\farcs7$ (254 km)} \\
02:18--03:15 & 3.52 & 46 & 1.045 & 0.075 & 53.4 & 44 & 15--500 & 0.46 & 27 & 1.11$\times$0.48 & 60$\times$26 \\ 
03:34--04:31 & 4.77 &  46 & 1.045 & 0.075 & 53.5 & 44 & 15--500 & 0.46 & 42 & 0.82$\times$0.49 & 44$\times$26 \\
04:50--05:47 & 6.10 &  46 & 1.045 & 0.075 & 53.6 & 44 & 15--500 & 0.47 &58 &  0.68$\times$0.49 & 37$\times$26 \\
06:05--07:02 & 7.27 &  46 & 1.045 & 0.075 & 53.7 & 44 & 15--500 & 0.47 & 73 & 0.63$\times$0.50 & 34$\times$27 \\
\hline
\multicolumn{11}{c}{Post-Impact ACA Observations, 2022 September 27, $\nu$ = 345 GHz, $\theta$\subs{MRS} = $19\farcs3$ (1050 km)} \\
02:33--03:36 & 3.77 &  45.7 & 1.045 & 0.075 & 53.4 & 9 & 9--49 & 0.46 & 32 &  2.95$\times$6.95 & 160$\times$378 \\
04:09--05:12 & 5.27 &  45.7 & 1.045 & 0.075 & 53.5 & 11 & 9--49 & 0.46 & 50 &  2.89$\times$4.99 & 157$\times$271 \\
05:44--06:48 & 7.02 &  45.7 & 1.045 & 0.075 & 53.6 & 11 & 9--49 & 0.47 & 69 & 2.79$\times$4.53 & 151$\times$246 \\
07:19--08:22 & 8.60 &  45.7 & 1.045 & 0.075 & 53.7 & 11 & 9--49 & 0.47 & 79 & 2.71$\times$4.39 & 147$\times$238 \\
\enddata
\tablecomments{\textit{T}\subs{int} is the total on-source integration time. \textit{r}\subs{H}, $\Delta$, and $\alpha$ are the heliocentric distance, geocentric distance, and solar phase angle (Sun--Asteroids--Earth), respectively, of Didymos-Dimorphos at the time of observations. $\nu$ is the mean frequency of each instrumental setting. \textit{N}\subs{ants} is the number of antennas utilized during each observation, with the range of baseline lengths indicated for each. PWV is the mean precipitable water vapor at zenith during the observations. $\theta$\subs{AR} is the angular resolution (synthesized beam) at $\nu$ given in arcseconds and in projected distance (km) at the geocentric distance of the asteroid system. $\theta$\subs{MRS} is the Maximum Recoverable Scale -- the largest scale on which the interferometer can recover flux -- at $\nu$ given in arcseconds and in projected distance (km) at the geocentric distance of the asteroid system.}
\end{deluxetable*}

\section{Results} \label{sec:results}
Our pre-impact observations sampled thermal continuum emission from the surfaces of Didymos-Dimorphos, and our post-impact observations simultaneously sampled thermal continuum emission from the asteroids and dust ejecta as well as spectral line emission from multiple species potentially present in the gas-phase ejecta. We detail our results for each epoch in turn.

\subsection{12 m Array Pre-Impact Observations} \label{subsec:pre-obs}
\begin{figure*}
\gridline{\fig{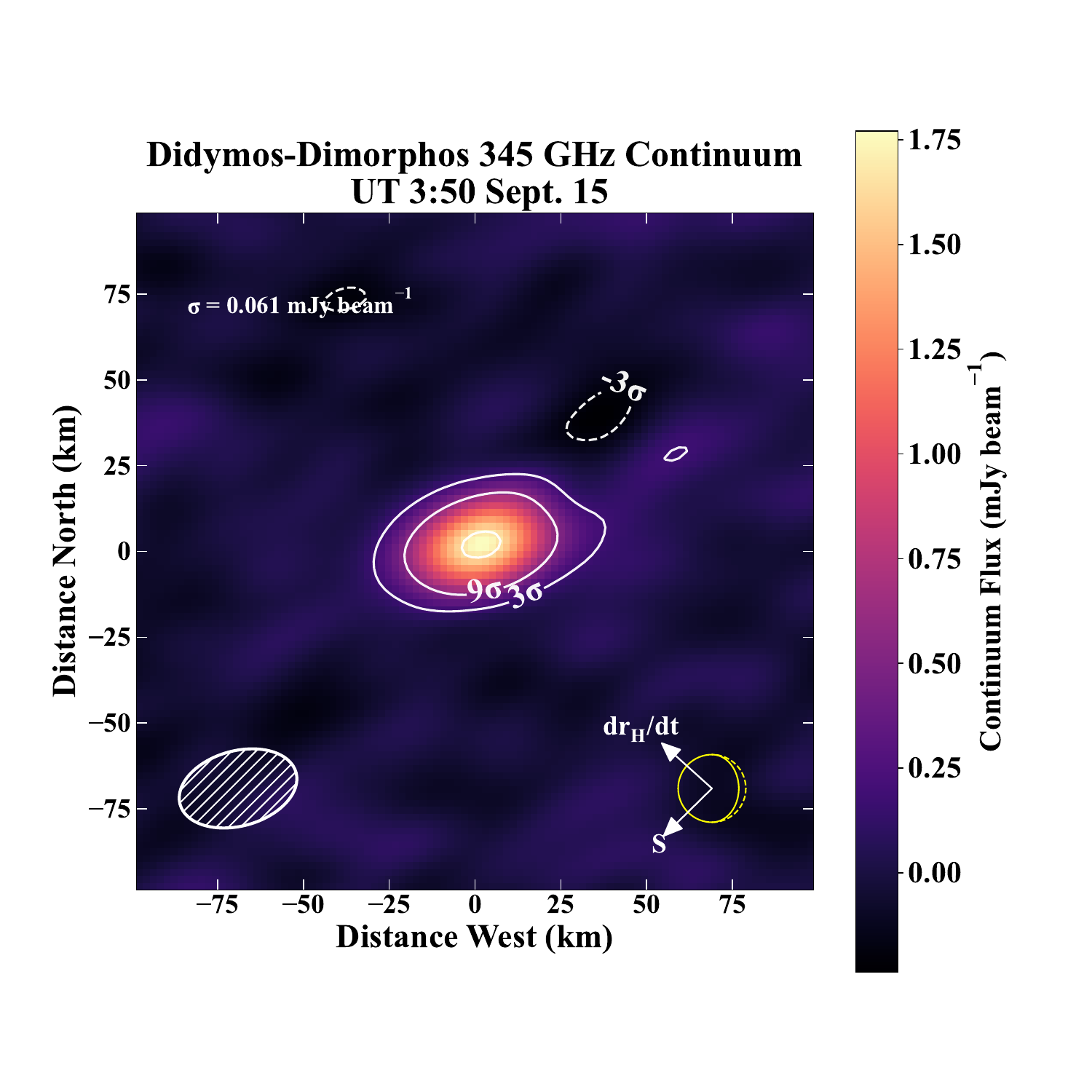}{0.50\textwidth}{(A)}
          \fig{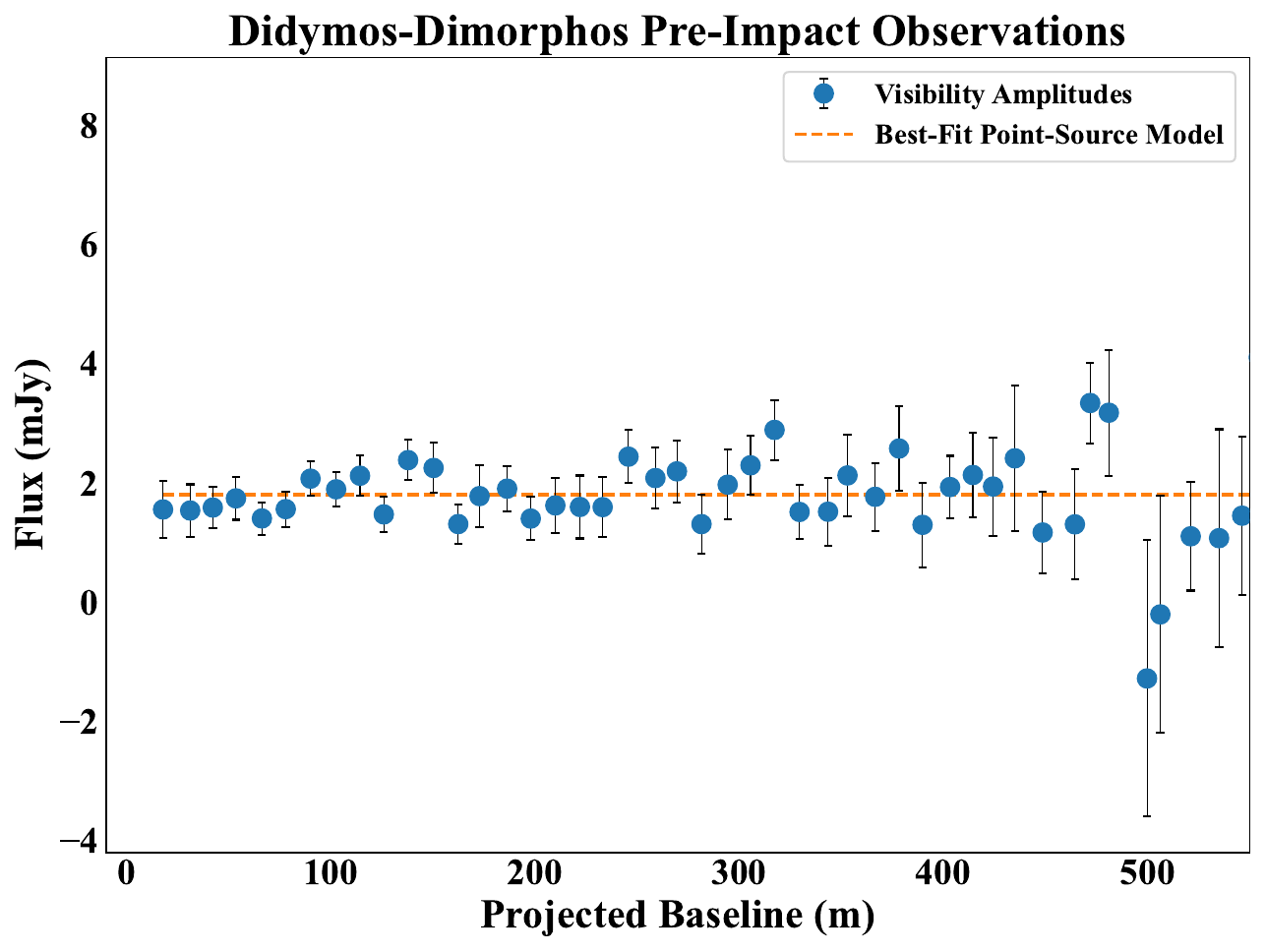}{0.50\textwidth}{(B)}
          }
\caption{\textbf{(A)} Pre-impact continuum flux map for Didymos-Dimorphos on UT 2022 September 15. The RMS noise ($\sigma$, mJy beam$^{-1}$) is indicated in the upper left corner. The contours are for the $3\sigma$, 9$\sigma$, and 27$\sigma$ levels. The size and orientation of the synthesized beam (0\farcs57$\times$0\farcs39) is indicated in the lower left corner. The asteroid's illumination phase ($\alpha \sim$ 38$\degr$), as well as the direction of the Sun (S) and the heliocentric velocity vector ($dr_H/dt$), are indicated in the lower right. (B) Pre-impact visibility amplitude vs.\ projected baseline for Didymos-Dimorphos (blue) as well as the best-fit CASA \texttt{uvmodelfit} point-source model (orange).
\label{fig:pre-impact}}
\end{figure*}

The high sensitivity of ALMA in TDM mode enabled us to measure Didymos-Dimorphos' thermal continuum emission in a short pre-impact execution of the 12 m array. Figure~\ref{fig:pre-impact} shows our pre-impact image of the asteroids alongside the measured visibility amplitudes vs.\ projected baseline. The image is consistent with an unresolved point source, as is the trend of constant visibility amplitude as a function of $uv$-distance. Using the CASA \texttt{uvmodelfit} routine with a point-source model returns a best-fit flux density of 1.79 $\pm$ 0.18 mJy and a reduced $\chi^2$ of 1.31, consistent with the flux density in the image (1.73 $\pm$ 0.18 mJy) and indicative of a good fit to the visibilities as a point-source.

\subsection{12 m Array Post-Impact Spectral Line Observations}\label{subsec:post-spec}
We did not detect spectral line emission from any of the targeted molecules in summed or individual executions, and here calculate 3$\sigma$ upper limits on the integrated intensity at the frequency of each transition. We assumed a line width of 2 \kms{}, consistent with the expansion velocity \vexp{} = 0.97 \kms{} measured for the fast-moving ejecta plume by \cite{Graykowski2023}, as well as the expansion velocity \vexp{} = 1.5--1.7 \kms{} measured for the alkali vapor plume detected by \cite{Shestakova2023}. Given the high expansion velocities for the fast ejecta and alkali vapor plumes (0.97--1.7 \kms{}), the gas-phase ejecta perpendicular to the line of sight would have been beyond the maximum recoverable scale (MRS; Table~\ref{tab:obslog}) of both the 12 m array and the ACA even during our first execution (T$+$3.52 hours) and filtered out of our images by the interferometer; thus, our upper limits are constraints on the gas-phase ejecta emission along the line of sight. \cite{Li2023} found that the axis of the ejecta plume was nearly parallel to the incoming direction of the DART spacecraft; thus, the mean motion of the ejecta compared to the LOS can be seen as opposite the direction of the DART spacecraft (``D'') in Figure~\ref{fig:post-impact1}.

We chose to calculate upper limits for the 12 m array owing to its higher sensitivity compared to the ACA.  We calculated 3$\sigma$ upper limits to the integrated emission intensity near the frequency of each transition within a 4\farcs68 diameter aperture centered on the asteroids (127 km radius from the asteroids, the MRS of the 12 m array). Our 3$\sigma$ upper limits for the integrated intensity of SiO, SiS, SO, AlO, KCl, and NaCN are $<2.1\times10^{-2}$ K \kms{}, $<1.5\times10^{-2}$ K \kms{}, $<1.3\times10^{-2}$ K \kms{}, $<1.7\times10^{-2}$ K \kms{}, $<1.3\times10^{-2}$ K \kms{}, and $<1.8\times10^{-2}$ K \kms{}, respectively.

\subsection{12 m Array and ACA Post-Impact Continuum Observations}\label{subsec:post-cont}
We examined the visibility amplitudes as a function of projected baseline averaged across all executions of the ACA and the 12 m array to characterize whether the ejecta were extended or point-like. For compact mm-sized ejecta with no resolved structure, we should expect to see a trend with constant amplitude in the visibilities, as in the pre-impact Figure~\ref{fig:pre-impact}B. Figures~\ref{fig:post-vis}A and~\ref{fig:post-vis}B demonstrate that instead the visibility amplitudes are consistent with an extended source, showing significant variation with baseline length, perhaps indicative of the complex, evolving ejecta structure imaged by HST \citep[see Figure 2 in][]{Li2023}. A full modeling treatment of the visibilities is beyond the scope of this manuscript and is the subject of future work.

\begin{figure*}
\gridline{\fig{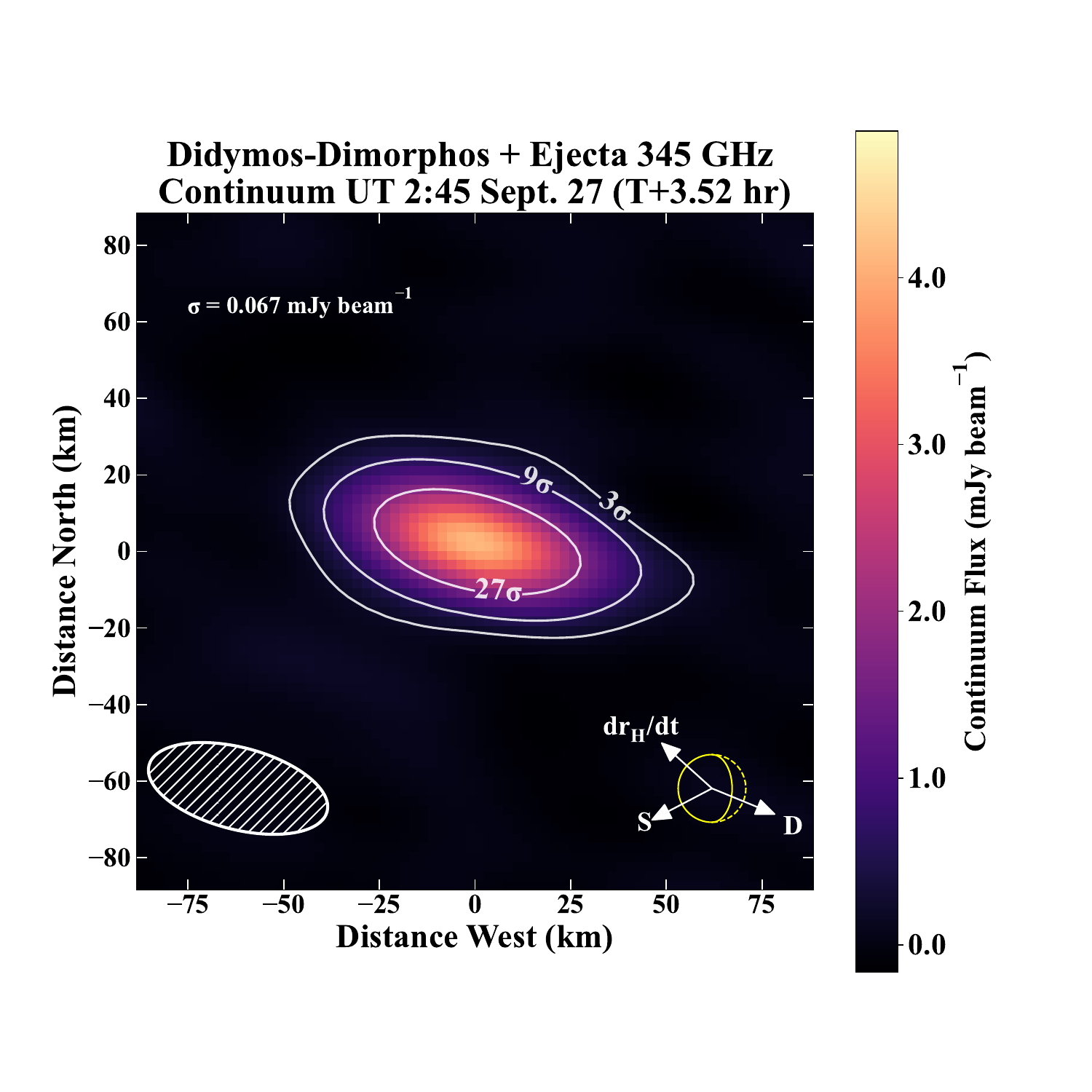}{0.50\textwidth}{(A)}
          \fig{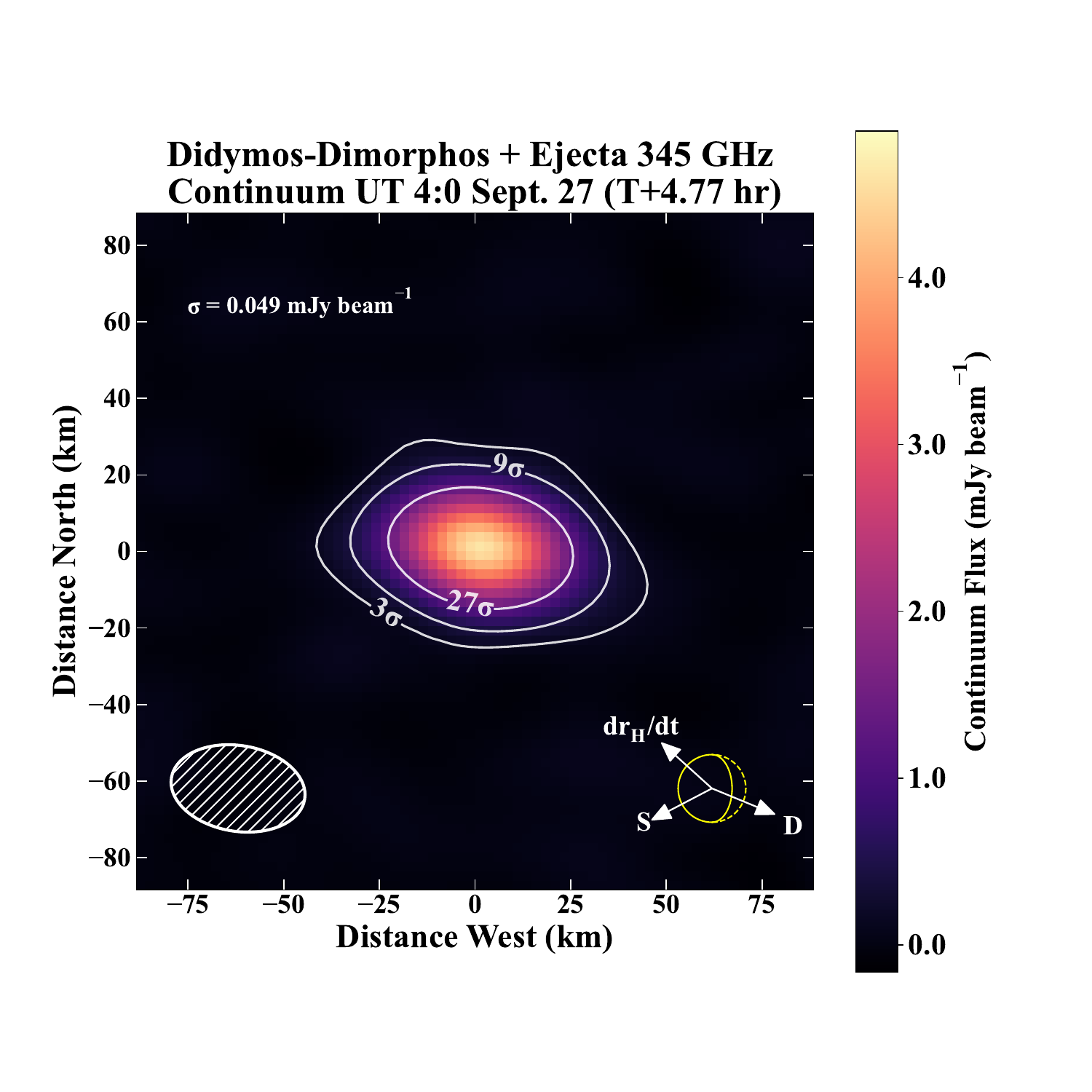}{0.50\textwidth}{(B)}
          }
\gridline{\fig{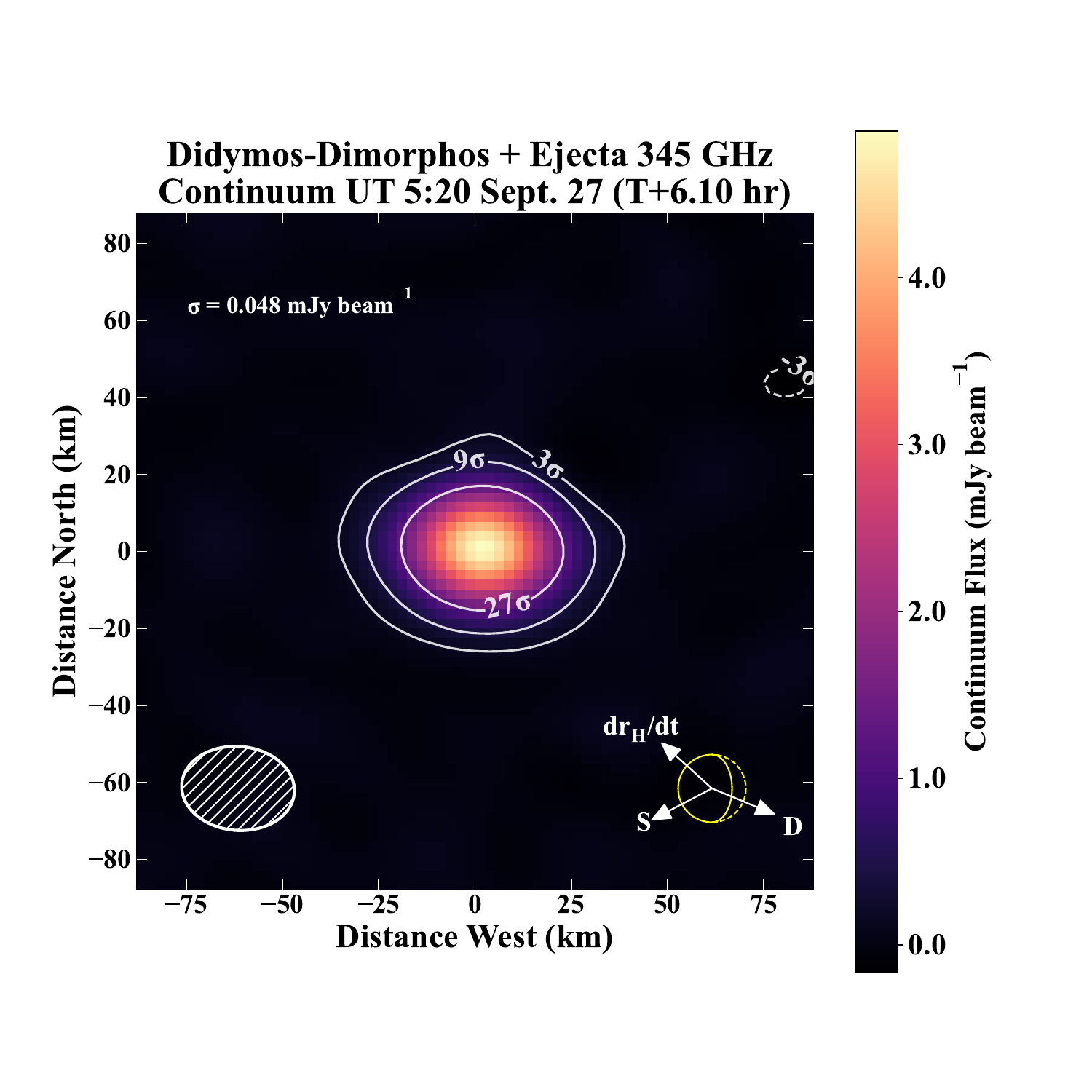}{0.50\textwidth}{(C)}
          \fig{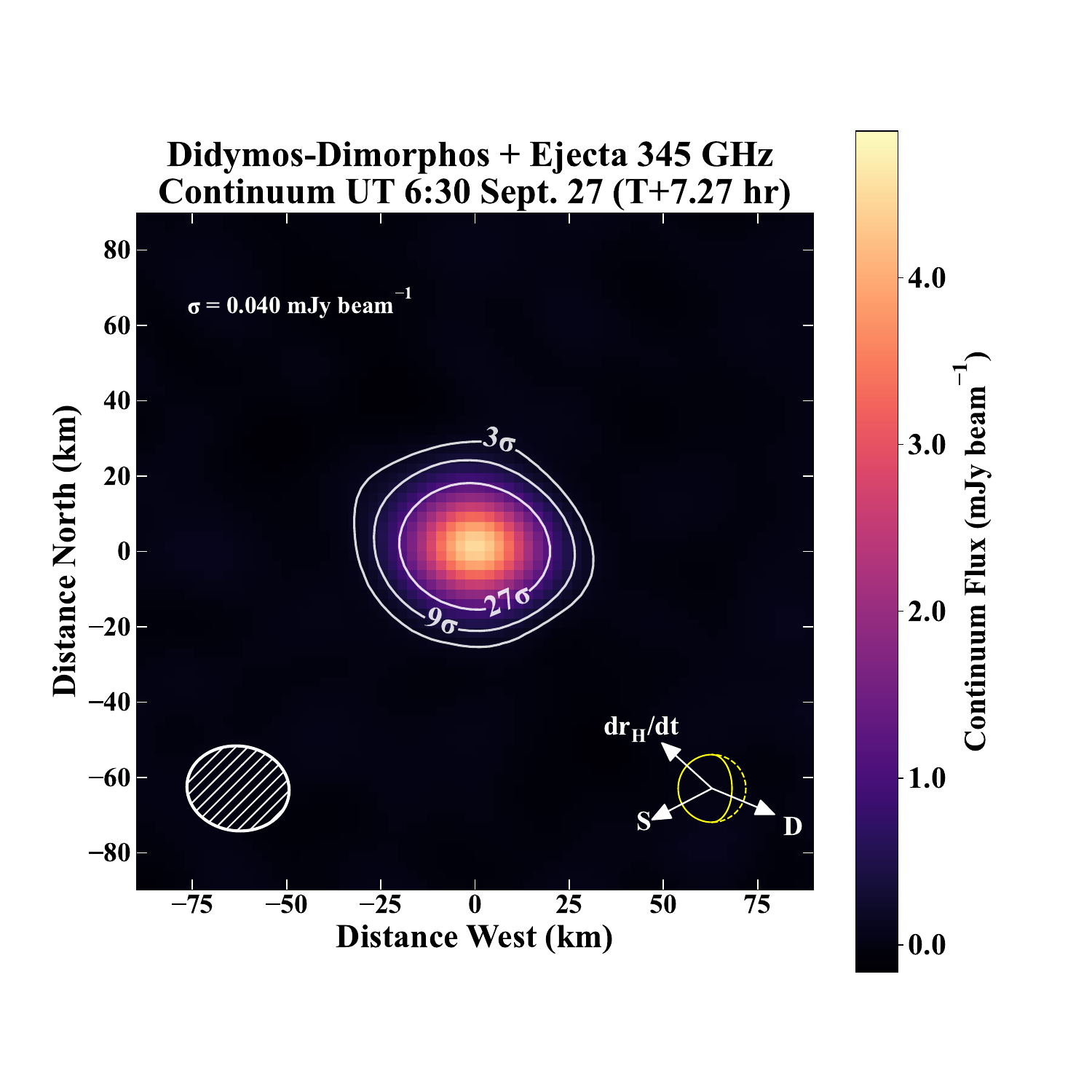}{0.50\textwidth}{(D)}
          }
\caption{\textbf{(A)--(D).} Post-impact continuum flux maps for Executions 1--4 imaged with the 12 m array, with traces and labels as in Figure~\ref{fig:pre-impact}. Arrows show the direction of the Sun (S), the DART spacecraft impact (D), and the asteroids' heliocentric velocity ($d\rh{}/dt$). Contours are for the 3$\sigma$, 9$\sigma$, and 27$\sigma$ levels.
\label{fig:post-impact1}}
\end{figure*}

\begin{figure*}
\gridline{\fig{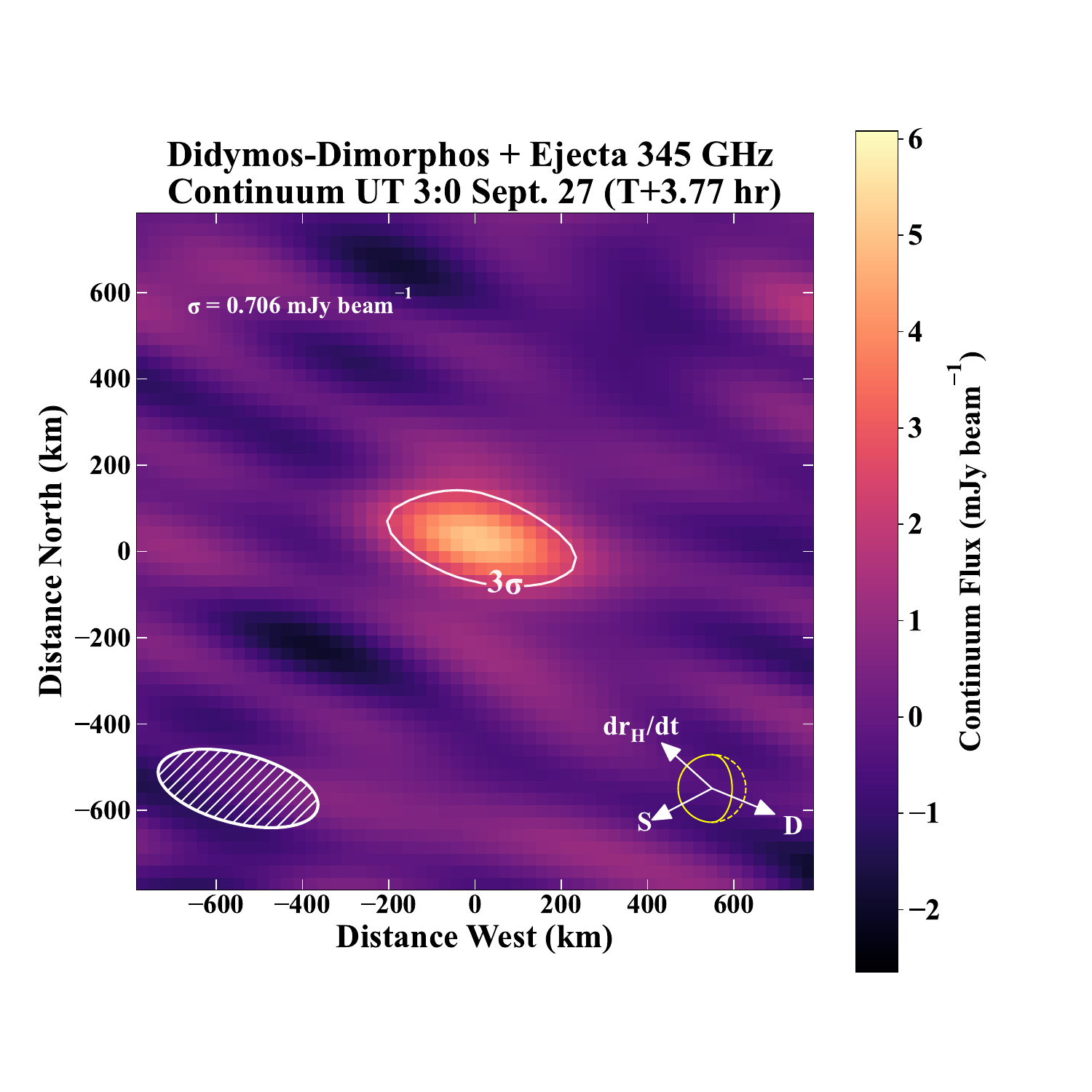}{0.50\textwidth}{(A)}
          \fig{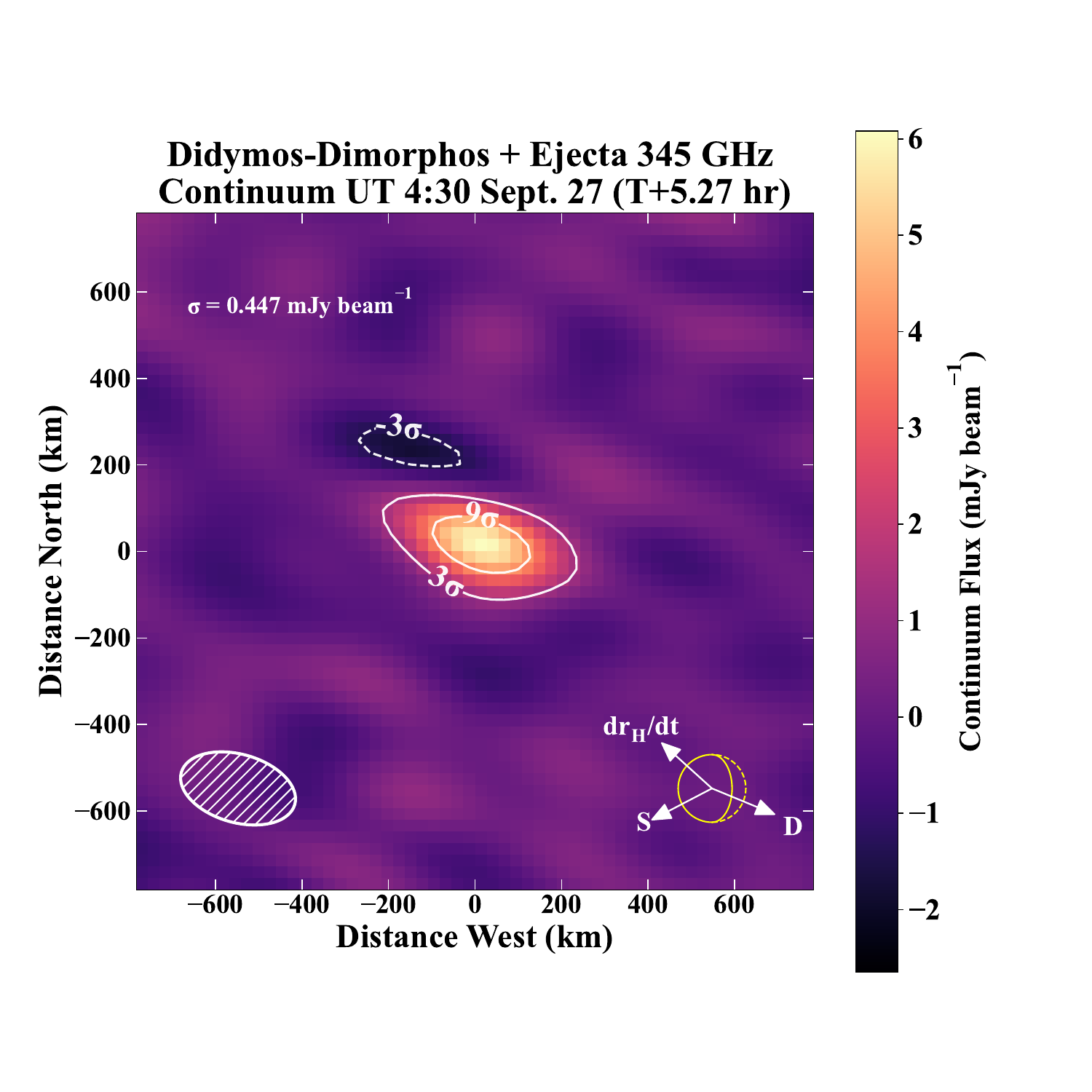}{0.50\textwidth}{(B)}
          }
\gridline{\fig{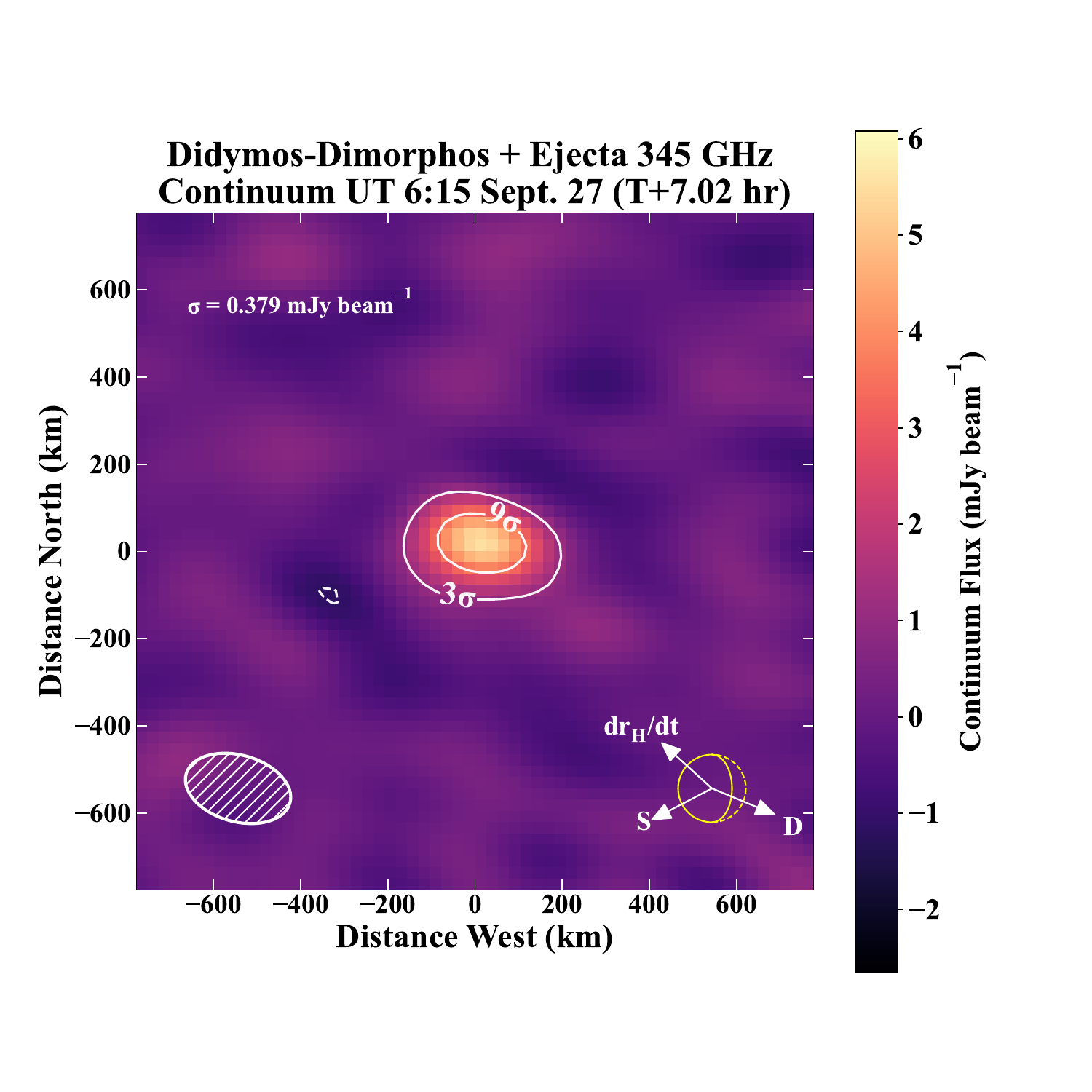}{0.50\textwidth}{(C)}
          \fig{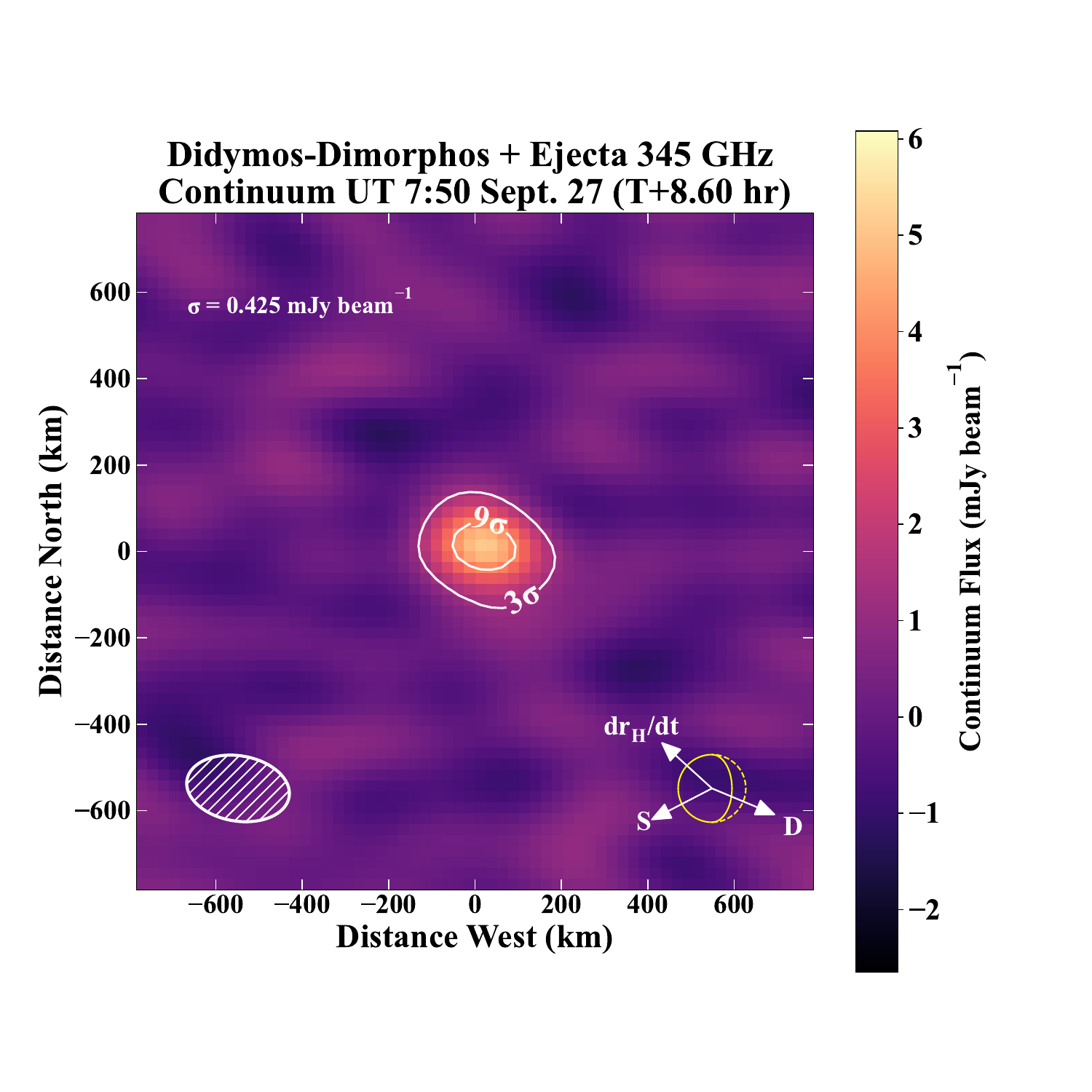}{0.50\textwidth}{(D)}
          }
\caption{\textbf{(A)--(B).} Post-impact continuum flux maps for Executions 1--4 imaged with the ACA, with traces and labels as in Figure~\ref{fig:post-impact1}.
\label{fig:post-impact2}}
\end{figure*}

\begin{figure*}
\gridline{\fig{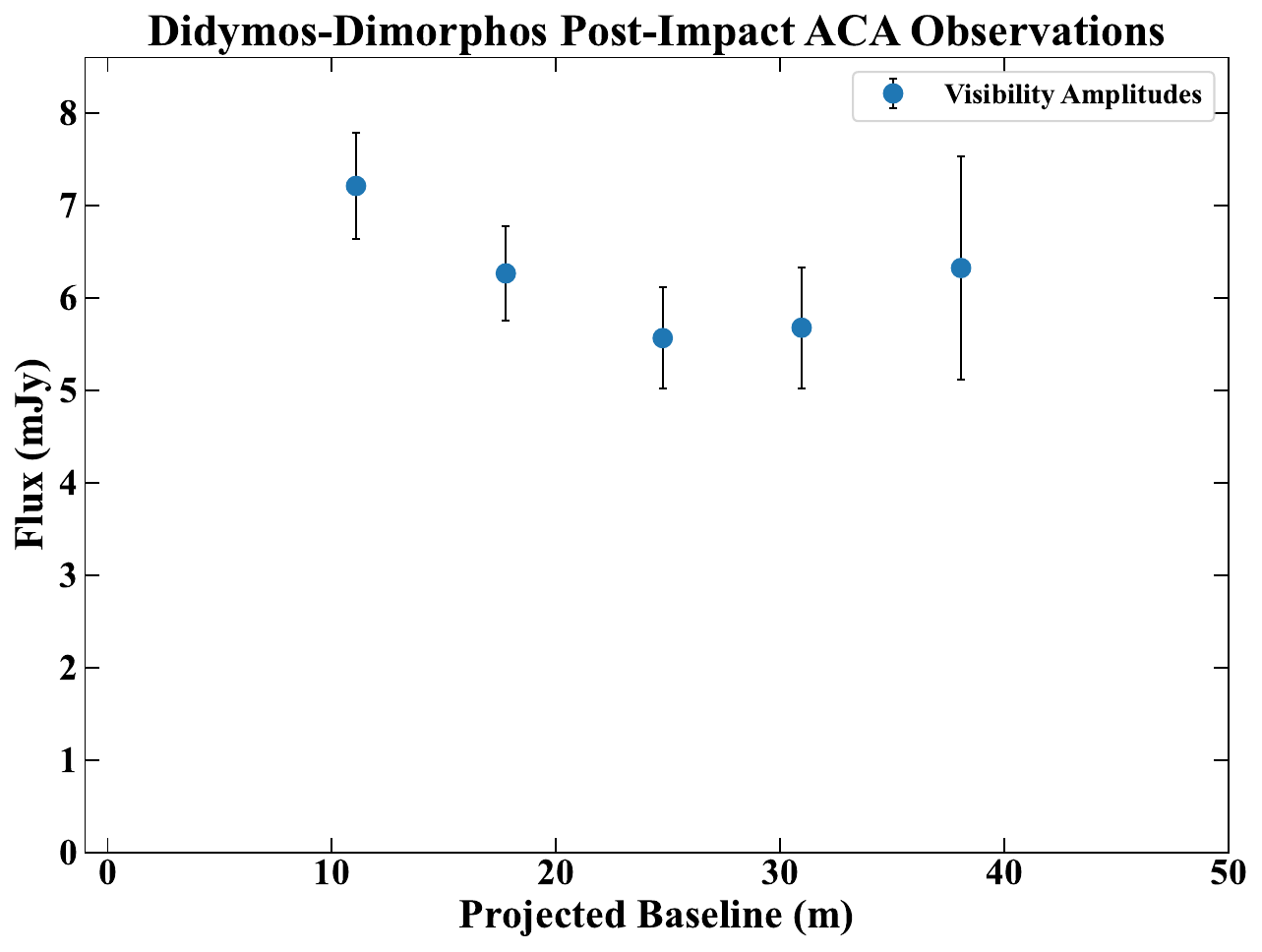}{0.45\textwidth}{(A)}
          \fig{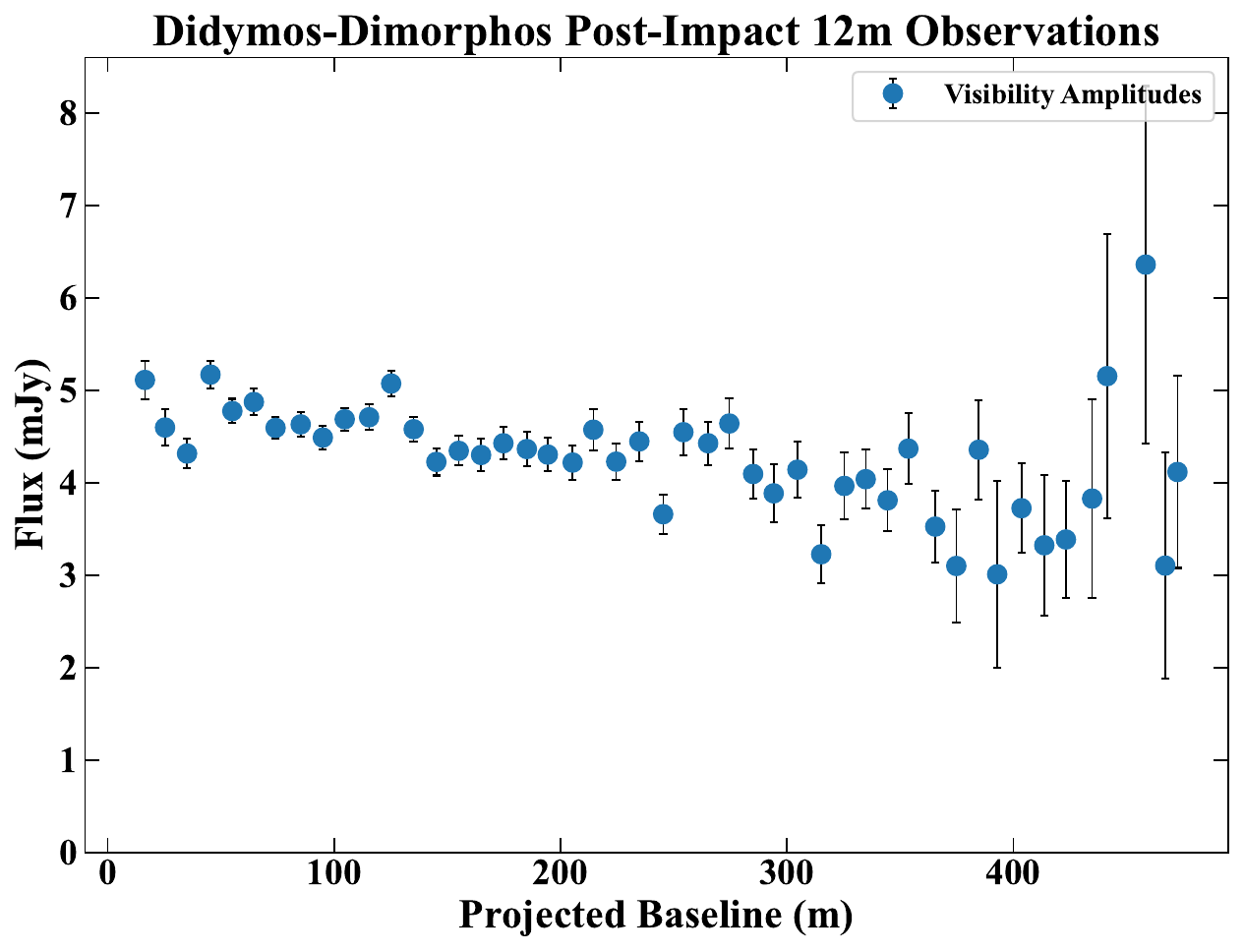}{0.45\textwidth}{(B)}
          }
\gridline{\fig{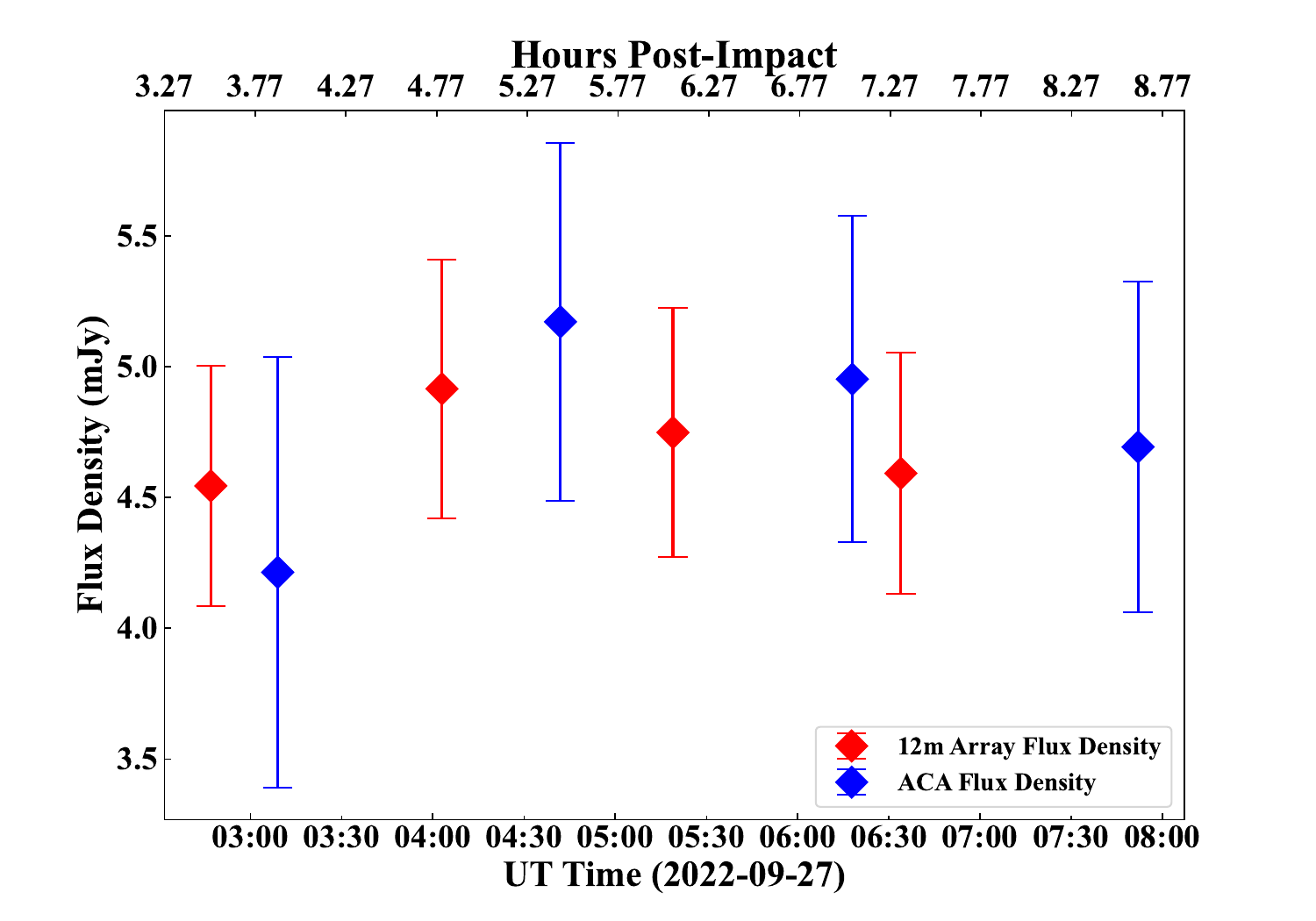}{0.55\textwidth}{(C)}
          }
\caption{\textbf{(A--B)} Post-impact visibility amplitude vs.\ projected baseline averaged for all ACA (A) and 12 m array (B) executions, demonstrating structure consistent with a complex extended source. (C) Post-impact ALMA lightcurve of Didymos-Dimorphos and the impact ejecta for each execution of the 12 m array. Flux densities were integrated within an aperture of 50 km radius for the 12 m array and of 200 km radius for the ACA. The estimated contribution of Didymos-Dimorphos is 2.74 mJy when neither are in eclipse (UT 02:18--05:45, 07:02--08:22) and 2.60 mJy during the secondary eclipse (UT 05:46 - 07:01).
\label{fig:post-vis}}
\end{figure*}

Our post-impact continuum images provide a record of the thermal emission from the asteroids and the dust ejecta from T$+$3.5 hours -- T$+$8.6 hours post-impact. Figures~\ref{fig:post-impact1} and~\ref{fig:post-impact2} show our post-impact continuum maps for the 12 m array and the ACA. Our images show extended ejecta, with statistically significant emission extending beyond the range of the synthesized beam; however, no clear structure (\eg{} jets) was found in the images. 

We integrated the flux inside an aperture of 50 km radius for the 12 m images and of 200 km radius for the ACA images, corresponding to radii which capture all emission at the $\geq3\sigma$ level. We provide a modeling treatment of the images in Section~\ref{sec:modeling} and list our measured fluxes there. Figure~\ref{fig:post-vis}B show that our measured flux densities are all in formal agreement, where the errorbars include both image RMS and the 10\% absolute flux calibration uncertainty.

\section{Modeling and Interpretation} \label{sec:modeling}

\subsection{Didymos-Dimorphos' Spectral Emissivity} \label{subsec:emissivity}
Our pre-impact observations sampled thermal emission from the surfaces of Didymos and Dimorphos, enabling us to measure Didymos-Dimorphos' spectral emissivity at millimeter wavelengths for the first time. Once their spectral emissivity is known, the flux attributable to Didymos-Dimorphos on the impact date can be calculated and subtracted from the measured post-impact fluxes to isolate that owing to the ejecta.

The Near Earth Asteroid Thermal model \citep[NEATM;][]{Harris1998,Delbo2002} was used to analyze thermal emission from the asteroids. The NEATM accounts for the solar phase angle ($\alpha$) during observations by considering the observed flux originating from the illuminated portion of the asteroid. We calculated the temperature at the sub-solar point ($T_{ss}$) as

\begin{equation}\label{eq:eq1}
T_{ss} = \left[\frac{(1-A)S_{\sun}}{\rh{}^2\epsilon\sigma\eta}\right]^{1/4}
\end{equation}

\noindent where A is the Bond albedo, S$_{\sun}$ is the Solar constant at 1 au (1360.8 W\,m$^{-2}$), \rh{} is the heliocentric distance (au), $\sigma$ is the Stefan Boltzmann constant $(5.67\times10^{-8}$ J\,s$^{-1}$\,m$^{-2}$\,K$^{-4}$), $\epsilon$ is the infrared emissivity \citep[assumed to be 0.9;][]{Delbo2004,Mainzer2016}, and $\eta$ the infrared beaming factor. We assumed a Bond albedo of 0.067 based on Didymos-Dimorphos' geometric albedo of 0.16 and a phase integral of 0.42 derived from the HG phase function using the DART Design Reference Asteroid parameters \citep{Rivkin2021,Bowell1989}, and a beaming parameter of 1.93 based on JWST observations of Didymos-Dimorphos taken at a similar solar phase angle to the ALMA observations \citep{Rivkin2023}. This is consistent with beaming parameters measured for other NEAs, which range from 0.5--2.5 in \textit{NEOWISE} surveys \citep{Mainzer2014,Mainzer2016}.  The resulting $T_{ss}$ is 324 K. The temperature was calculated across the asteroid surface assuming that no emission originates from the night side as

\begin{equation}\label{eq:eq2}
T(\theta,\phi) = T_{ss}(\cos\phi\cos\theta)^{1/4}
\end{equation}

\noindent for $\theta \in [\alpha-\pi/2,\alpha+\pi/2]$ and $T(\theta,\phi) = 0$ elsewhere. The expected flux ($F_{\nu}$) at a frequency $\nu$ was then calculated as

\begin{equation}\label{eq:eq3}
F_\nu = \frac{\epsilon(\nu)d^2}{\Delta^2}\frac{h\nu^3}{c^2} \int_0^{\pi/2} \int_{-\pi/2}^{\pi/2} \frac{\cos^2\phi\cos(\theta-\alpha)}{\exp\left( \frac{h\nu}{k_BT(\theta,\phi)} \right)} \mathrm{d}\theta\mathrm{d}\phi
\end{equation}

\noindent where $\epsilon(\nu)$ is the frequency-dependent spectral emissivity, $k_B$ is the Boltzmann constant ($1.38\times10^{-23}$ J\,K$^{-1}$), $d$ is the asteroid diameter (m), and $\Delta$ the geocentric distance (m). As the asteroids were not spatially resolved during our pre-impact observations, we assumed that they have the same spectral emissivity. Using the best-fit flux from our point-source model fit (1.79 $\pm$ 0.18 mJy) and the diameters derived by \cite{Daly2023}, we calculated Didymos' and Dimorphos' contributions to the above integral and find an emissivity of $\epsilon(\nu)$ = 0.95 $\pm$ 0.10 at $\nu$ = 343.5 GHz. Incorporating our millimeter emissivity into thermophysical modeling of Didymos-Dimorphos alongside the JWST measurements \citep{Rivkin2023} is the subject of future work.

Previous observations of asteroids at millimeter wavelengths include ALMA observations of 1 Ceres \citep{Li2020,Li2022}, 3 Juno \citep{ALMA2015}, and 16 Psyche \citep{deKleer2021}, Rosetta MIRO observations of 21 Lutetia and 2867 Steins \citep{Gulkis2010,Gulkis2012}, and South Pole Telescope detections of five other main-belt asteroids \citep{Chichura2022}. ALMA imaged Ceres around 265 GHz and Juno and Psyche between 223 and 243 GHz.  MIRO observed Lutetia and Steins near 190 and 560 GHz.  The South Pole Telescope detected 13 Egeria and 22 Kalliope at 150 GHz; 324 Bamberga between 95 and 150 GHz; and 772 Tanete and 1093 Freda and between 95 GHz and 215 GHz.

These main-belt asteroids show an apparent correlation between average millimeter emissivity and spectral classification / surface composition.  Observed C-class objects (Ceres, Bamberga, Egeria) and S-class objects (Juno) have millimeter emissivity 0.8--1.0.  Observed M-class objects have a wider range of millimeter emissivity, from near 1 (Lutetia) down to $\sim$0.6 (Psyche, Kalliope); while the E-class Steins has emissivity 0.6 to 0.9 depending on wavelength.  Low emissivity for Psyche and Kalliope is interpreted as due to high surface metal content \citep{deKleer2021}. Our observed average emissivity of 0.95 ± 0.10 for Didymos-Dimorphos is consistent with its S-class classification and a silicate surface composition with minimal surface metal.

With the millimeter spectral emissivity known, we repeated our calculations for Didymos-Dimorphos on the impact date at \rh{} = 1.045 au and $\Delta$ = 0.075 au, where we find $T_{ss}$ = 330 K. We predict a flux of 2.74 mJy attributable to the asteroids themselves when neither are in eclipse, which is applicable to Executions 1, 2, and 3 of the 12 m array and Executions 1, 2, and 4 of the ACA. \cite{Scheirich2022} developed a photometric model for predicting mutual events for Didymos-Dimorphos, with secondary eclipses resulting in a 0.049 mag decrease in brightness. We reduced our predicted Didymos-Dimorphos flux a corresponding amount to 2.60 mJy for comparison with our observations during secondary eclipse (Execution 4 of the 12 m array and Execution 3 of the ACA).

\subsection{Analysis of the Ejecta}\label{subsec:ejecta}
Interferometers act as spatial filters owing to the lack of $uv$-coverage for the smallest spatial frequencies (i.e., two antennas can only be placed so physically close to one another, resulting in a lack of very short baselines). The maximum recoverable scale (MRS) is a characteristic spatial scale for a given interferometric array configuration, and flux on angular scales larger than the MRS cannot be recovered. For our post-impact measurements with the 12 m array at 345 GHz in configuration C3, the MRS is 4\farcs68, corresponding to 254 km at the geocentric distance of the asteroids ($\Delta$ = 0.075 au). For the ACA the $19\farcs3$ MRS ($\sim$1050 km) is much larger, although the sensitivity of 11 $\times$ 7 m antennas is lower than that of the 44 $\times$ 12 m antennas of the main array. 

HST imaging of the DART impact showed a complex, multi-featured ejecta structure evolving throughout the time of our ALMA observations, with some features stretching hundreds of km from the asteroids \citep{Li2023}. Although HST is sensitive to smaller particle sizes than ALMA, this places the scale of the ejecta into context with our observations and the MRS of the 12 m array and the ACA. Thus, our measurements are lower limits on the total flux at sub-millimeter wavelengths, recovering flux from particles within 127 km and 525 km radii from Didymos-Dimporphos for the 12 m array and the ACA, respectively. 

We calculated the ejecta flux density for each execution by subtracting our predicted flux density for Didymos-Dimorphos, using the appropriate value depending on whether the asteroids were in eclipse. We incorporated an additional 10\% uncertainty into our ejecta flux densities to account for systematic uncertainties in our prediction of Didymos-Dimorphos' post-impact flux. We used these derived ejecta flux densities to calculate the ejecta mass responsible for the flux density measured in each execution.

\subsubsection{Calculation of the Dust Ejecta Masses}\label{subsubsec:masses}
Dust masses were calculated for each execution following the methods of \cite{Jewitt1990} and \cite{Boissier2012}. We calculated the grain absorption efficiency, $Q$\subs{abs}, at $\lambda=0.87$ mm using Mie theory for spherical, homogeneous grains with the \texttt{miepython} package\footnote{The software is available on GitHub \texttt{miepython} codebase: \url{https://github.com/scottprahl/miepython} under an MIT license \citep{Prahl2022}.}. Calculating $Q$\subs{abs} requires knowledge of the grain size and the complex refractive index, $m_\lambda$, which is directly related to the optical constants $(n,k)$ of the material as $m_\lambda=n-ik$. The composition and optical properties of ordinary chondrites are likely the best proxy to that of the ejecta, as sample return material from the Hayabusa mission to asteroid Itokawa \citep{Nakamura2011} demonstrated a link between S-type asteroids such as Didymos-Dimorphos \citep{deLeon2006,Cheng2018} and ordinary chondrites. We are not aware of optical constants at $\lambda = 0.87$ mm measured for ordinary chondrite particles in the literature, and there is a general paucity of optical constants at $\lambda = 0.87$ mm for materials that may be similar to the S-type regolith within Dimorphos. A mixture of crystalline olivine and crystalline pyroxene may serve as a first-order approximation to ordinary chondrite material \citep{Lawrence2007} and thus representative of the grain composition. Although there are optical constants reported for low-Fe crystalline olivines \citep{Fabian2001} and for amorphous olivines and pyroxenes \citep{Jager2003}, as far as we are aware, there are no available optical constants for crystalline pyroxenes at $\lambda = 0.87$ mm.

\begin{figure*}
\plotone{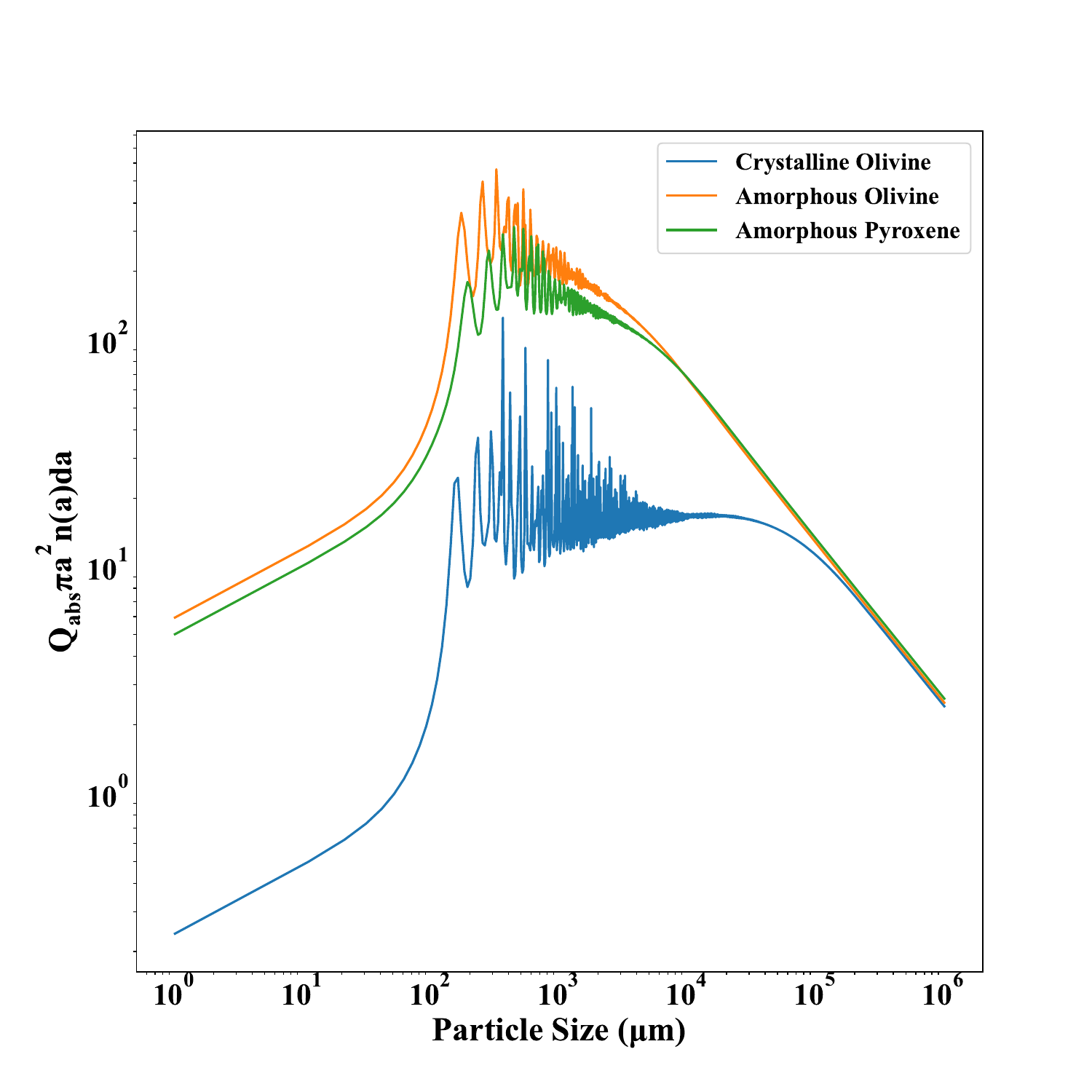}
\caption{Ejecta absorption cross section $Q_{abs}\pi a^2 n(a)\,da$ at $\lambda=0.87$ mm using a particle distribution of $a^{-2.7}$ for crystalline olivine, amorphous olivine, and amorphous pyroxene. We used a piece-wise differential particle size distribution of $a^{-2.7}$ for particles of radius 0.5 $\mu$m to 2 mm and of $a^{-3.7}$ for larger particles up to a radius of 2.5 cm \citep{Li2023} when calculating ejecta masses. \label{fig:plotqabs}}
\end{figure*}

In light of this situation, we used optical constants to calculate refractive indexes for three separate materials when calculating $Q$\subs{abs}: (1) low-Fe crystalline olivine (Mg$_{1.9}$Fe$_{0.1}$SiO$_4$), (2) amorphous olivine (Mg$_2$SiO$_4$), and (3) amorphous pyroxene (MgSiO$_3$). We then considered the grain porosity for ordinary chondrites. \cite{Britt2003} found a porosity of 6\% for H and L chondrites and 9\% for LL chondrites. We assumed an 8.2\% average porosity of ordinary chondrites \citep{Tsuchiyama2011}. The final refractive index for the porous grains was calculated using the two-component Maxwell-Garnett formula assuming vacuum as the core and the minerals as inclusions in the matrix (\ie{} a hollow sphere). We note that EMT theories only provide approximations to the real optical properties of a substance, and that the results are most accurate when the grain size is small compared to the wavelength and the inclusions small compared to the matrix \citep{Ossenkopf1991}. More in-depth theories are beyond the scope of this manuscript. Figure~\ref{fig:plotqabs} shows calculated absorption cross sections as a function of particle size for the materials considered in this work.

Assuming the dust size distribution and chemical properties do not vary within the field of view, the flux density is given as \citep{Jewitt1990}:

\begin{equation}\label{eq:eq4}
S_\lambda = \frac{2k_B}{\lambda^2\Delta^2}\int_{a_{min}}^{a_{max}}T(a)Q_{abs}\pi a^2n(a)\,\mathrm{d}a
\end{equation}

where $n(a)\,\mathrm{d}a\propto a^q$ is the differential particle size distribution, $q$ is the size index, $a_{min}$ and $a_{max}$ are the minimum and maximum grain radii, $T(a)$ is the temperature of the grains, and $Q_{abs}$ is the grain absorption efficiency factor. The grain temperature is assumed to be independent of size and equal to the blackbody equilibrium temperature for fast-rotating bodies of 266 K at \rh{} = 1.045 au. Applying the Rayleigh-Jeans approximation for blackbody emission, the flux density can be related to the dust opacity $\kappa_\lambda$ as \citep{Jewitt1992,Boissier2012}:

\begin{equation}\label{eq:eq5}
S_\lambda = \frac{2k_BTM\kappa_\lambda}{\lambda^2\Delta^2}
\end{equation}

\noindent where the dust opacity is defined as:

\begin{equation}\label{eq:eq6}
\kappa_\lambda = \frac{\int_{a_{min}}^{a_{max}}Q_{abs}\pi a^2n(a)\,da}{\int_{a_{min}}^{a_{max}}(4\pi/3)\rho a^3n(a)\,\mathrm{d}a}
\end{equation}

\noindent where $\rho$ is the grain density. The dust opacity (m$^2$\,kg$^{-1}$) expresses the effective surface area for absorption available per unit mass and is related to the optical depth of the material. We assumed $\rho = 3500$ kg\,m$^{-3}$ as measured in ordinary chondrites and the Itokawa returned samples \citep{Tsuchiyama2011}. We used a a piece-wise differential particle size distribution of $a^{-2.7}$ for particles of radius 0.5 $\mu$m to 2 mm and of $a^{-3.7}$ for larger particles up to a radius of 2.5 cm as derived by \cite{Li2023} based on HST imaging of the ejecta.

\begin{deluxetable*}{ccccc}
\tablenum{3}
\tablecaption{Refractive Indexes $m_\lambda = n-ik$ and Dust Opacities $\kappa_\lambda$ at $\lambda = 0.87$ mm \label{tab:constants}}
\tablewidth{0pt}
\tablehead{
\colhead{ID} & \colhead{Material} & \colhead{$n$} & \colhead{$k$} & \colhead{$\kappa_\lambda$} \\
\colhead{} & \colhead{} & \colhead{} & \colhead{} & \colhead{(m$^2$\,kg$^{-1}$)} 
}
\startdata
Cry-Oli & Crystalline olivine (Mg$_{1.9}$Fe$_{0.1}$SiO$_4)$\sups{(a)} &  2.60 & $8.19\times10^{-4}$ & $4.15\times10^{-3}$\\
Amo-Oli & Amorphous olivine (Mg$_2$SiO$_4$)\sups{(b)} &  2.35 & $1.66\times10^{-2}$ & $2.05\times10^{-2}$\\
Amo-Pyr & Amorphous pyroxene (MgSiO$_3$)\sups{(b)} &  2.07 & $1.11\times10^{-2}$ & $1.88\times10^{-2}$\\
\enddata
\tablecomments{\sups{(a)}Optical constants from \cite{Fabian2001}. \sups{(b)}Optical constants from \cite{Jager2003}. In all cases the dust opacities were computed in this work assuming minimum and maximum particles of radius 0.5 $\mu$m and 2.5 cm, respectively, with a piece-wise differential particle size distribution of $a^{-2.7}$ for particles of radius 0.5 $\mu$m to 2 mm and of $a^{-3.7}$ for larger particles up to a radius of 2.5 cm \citep{Li2023}. Abbreviations/IDs are given as referenced in Table~\ref{tab:masses} for each material.
}
\end{deluxetable*}

We calculated $\kappa_\lambda$ at $\lambda = 0.87$ mm for crystalline olivine, amorphous olivine, and amorphous pyroxene assuming a minimum particle radius of 0.5 $\mu$m and maximum particle radius of 2.5 cm based on the size distribution measured by \cite{Li2023}. Figure~\ref{fig:plotqabs} demonstrates that the ejecta absorption cross section (and thus the flux density) is dominated by particles in the sub-mm to mm size range. Using Equation~\ref{eq:eq4} and our calculated $Q_{abs}$, particles with radii between 100 $\mu$m and 10 mm contribute $\sim$75\% of our measured flux density at $\lambda$ = 0.87 mm. Table~\ref{tab:constants} lists the optical constants and dust opacities used in this work, and Table~\ref{tab:masses} lists the measured flux density, derived ejecta flux density, and calculated ejecta mass for each mineral composition considered.

\begin{deluxetable*}{cccccc}
\tablenum{4}
\tablecaption{Ejecta Flux Densities and Mass Estimates\label{tab:masses}}
\tablewidth{0pt}
\tablehead{
\colhead{Time\sups{(a)}} & \colhead{Total Flux\sups{(b)}} & \colhead{Ejecta Flux\sups{c}} & \multicolumn{3}{c}{Ejecta Mass ($10^7$ kg)\sups{(d)}}  \\ \cline{4-6}
\colhead{(Post-Impact)} & \colhead{(mJy)} & \colhead{(mJy)} & \colhead{Cry-Oli} & \colhead{Amo-Oli} & \colhead{Amo-Pyr} 
}
\startdata
3.52$^\dag$  & 4.54 $\pm$ 0.46 & 1.80 $\pm$ 0.53 & 5.67 $\pm$ 1.68 & 1.15 $\pm$ 0.34 & 1.25 $\pm$ 0.37 \\
3.77$^\ddag$ & 4.21 $\pm$ 0.82 & 1.47 $\pm$ 0.89 & 4.63 $\pm$ 2.72 & 0.94 $\pm$ 0.55 & 1.02 $\pm$ 0.60 \\
4.77$^\dag$ &  4.92 $\pm$ 0.49 & 2.17 $\pm$ 0.56 & 6.84 $\pm$ 1.55 & 1.39 $\pm$ 0.31 & 1.51 $\pm$ 0.34  \\
5.27$^\ddag$ & 5.17 $\pm$ 0.68 & 2.43 $\pm$ 0.74 & 7.65 $\pm$ 2.32 & 1.55 $\pm$ 0.47 & 1.69 $\pm$ 0.51  \\
6.10$^\dag$  & 4.75 $\pm$ 0.48 & 2.01 $\pm$ 0.56 & 6.31 $\pm$ 1.50 & 1.28 $\pm$ 0.30 & 1.39 $\pm$ 0.33  \\
7.02$^\ddag$ & 4.95 $\pm$ 0.62 & 2.35 $\pm$ 0.68 & 7.39 $\pm$ 2.13 & 1.50 $\pm$ 0.43 & 1.63 $\pm$ 0.47 \\
7.27$^\dag$  & 4.59 $\pm$ 0.46 & 1.99 $\pm$ 0.54 & 6.26 $\pm$ 1.45 & 1.27 $\pm$ 0.29 & 1.38 $\pm$ 0.32  \\
8.60$^\ddag$ & 4.69 $\pm$ 0.63 & 1.95 $\pm$ 0.69 & 6.14 $\pm$ 2.17 & 1.24 $\pm$ 0.44 & 1.36 $\pm$ 0.48  \\
\hline
Weighted Average\sups{(e)} & 4.72 $\pm$ 0.19 & 2.03 $\pm$ 0.22 & 6.39 $\pm$ 0.64 & 1.29 $\pm$ 0.13 & 1.41 $\pm$ 0.14 \\
\enddata
\tablecomments{\sups{(a)}Time post-impact (hours). $^\dag$12 m execution. $^\ddag$ACA execution. \sups{b}Flux density for Didymos-Dimorphos plus the ejecta integrated in an aperture of 50 km radius for 12m array executions and of 200 km radius for the ACA. Uncertainties include image RMS plus a 10\% uncertainty in absolute flux calibration. \sups{c}Flux density after subtracting the 2.74 mJy or 2.60 mJy flux density of Didymos-Dimorphos outside or during secondary eclipse, respectively.} Uncertainties include an additional 10\% to account for the subtraction of Didymos-Dimorphos' calculated flux.  \sups{d}Dust mass calculated following the methods of \cite{Jewitt1990} and \cite{Boissier2012} as outlined in Section~\ref{subsubsec:masses}. Masses are given assuming grain compositions of crystalline olivine, amorphous olivine, or pure amorphous pyroxene. \sups{(e)}Weighted average of all flux  and mass measurements.
\end{deluxetable*}

These ALMA measurements provide an independent constraint on the mass of ejecta present within the MRS (127 km and 525 km radii from the asteroids for the 12 m array and ACA, respectively) between T$+$3.52 and T$+$8.60 hours post-impact. This timeframe corresponds to the earliest stages of the ejecta evolution, before complicated dynamical interactions between the asteroids and the ejecta began and before the ejecta tail had fully formed \citep{Li2023,Graykowski2023}.

The measured fluxes are in formal agreement between arrays and between executions. The slight decrease in flux during 12 m Execution 4 and ACA Execution 3 (Table~\ref{tab:masses}, Figure~\ref{fig:post-vis}B) corresponds with a secondary eclipse. Given the overall good agreement between the fluxes, we calculated the weighted average flux of all executions to compare with results for individual executions. We consider this weighted average to be representative of the dust mass lower limit for the ejecta.

By examining the trends in our absorption efficiencies (Figure~\ref{fig:plotqabs}) and ejecta dust masses (Table~\ref{tab:masses}), we can gain insight into how each considered material affected our calculations. The imaginary part of the refractive index for low-Fe crystalline olivine at $\lambda$ = 0.87 mm is the smallest among the materials in this work, and stands out as presenting the lowest absorption efficiency $Q_{abs}$. For the range in particle radii (0.5 $\mu$m--2.5 cm) considered here, low-Fe crystalline olivine grains provide the highest ejecta masses of $\sim6.4\times10^7$ kg, whereas amorphous olivine provides some of the lowest ($\sim1.3\times10^7$ kg). Given these trends and that crystalline material is likely more prevalent than amorphous in ordinary chondritic material (such as that contained in Dimorphos' grains), the higher mass derived from crystalline olivine is favored.

Given the presence of meter-sized boulders in the final images of Dimorphos taken by the DART spacecraft before impact \citep{Daly2023}, it is possible that particles up to meters in size were present in the ejecta. Unfortunately, the differential particle size distribution measured by HST \citep{Li2023} is only valid for particles up to a few cm, and the size distribution for meter-sized material in the ejecta is currently unknown. If we assume that the meter-sized material followed the same particle size distribution as the cm-sized particles in the HST imaging ($n(a)\propto a^{-3.7}$) and a maximum particle radius $a_{max}$ = 0.5 m, our resulting ejecta masses are higher by roughly a factor of two. However, our assumptions regarding the particle size distribution for meter-sized material are a significant caveat, and a firm measure of the ejecta mass that incorporates meter-sized material will require input from long-term follow up observations of the ejecta tail as well as dynamical simulations of the ejecta.

The range of ejecta masses calculated based on our weighted average flux measurements (1.3--$6.4\times10^7$ kg depending on the assumed grain properties) represents 0.3--1.5\% of Dimorphos' total mass of $4.3\times10^9$ kg \citep{Daly2023} and is consistent with pre-impact modeling estimates \citep{Raducan2022a,Stickle2022}. Our calculated ejecta masses are consistent with lower limits of 0.3--0.5\% of Dimorphos' mass estimated by \cite{Graykowski2023} based on optical wavelength measurements of the ejecta. \cite{Moreno2023} conducted optical imaging and photometry, and performed an independent analysis of the HST imaging presented in \cite{Li2023}. Their lower limits on the ejecta mass of $>$4--5$\times10^6$ kg are consistent with our results.

\subsubsection{Upper Limits on Gas-Phase Ejecta}\label{subsubsec:gasphase}
Using our calculated $3\sigma$ upper limits on integrated intensity for our targeted spectral line transitions (Section~\ref{subsec:post-spec}), we derived upper limits on column densities and masses of gas-phase ejecta along the line-of-sight during our observations. The column density of the gas-phase ejecta within the upper transition state ($<N_u>$) can be related to the integrated intensity as \citep{Bockelee1994}:

\begin{equation}\label{eq:eq7}
<N_u> = \frac{8 \pi \nu^2 k_B}{h c^3 A_{ul}}\int T_b\,dv
\end{equation}

\noindent where $\nu$ is the line frequency, $k_B$ is the Boltzmann constant, and $A_{ul}$ the Einstein-A coefficient. The molecular column density in the beam $<N>$ is then related to $<N_u>$ as:

\begin{equation}\label{eq:eq2}
<N_u> = \frac{<N>g_u}{Z(T_{rot})}e^{-E_u/k_BT_{rot}}
\end{equation}

\noindent where $g_u$ and $E_u$ are the statistical weight and energy of the upper transition state, respectively, and $Z(T_{rot})$ is the rotational partition function evaluated at $T_{rot}$. We assumed a gas rotational temperature $T_{rot}$ = 1000 K based on pre-impact modeling of temperatures achieved in the ejecta \citep{Raducan2022b}. We used rotational partition functions, $Z(T_{rot}$ = 1000 K), reported in CDMS for SiO, SiS, AlO, and KCl \citep{Endres2016}, in HITRAN for SO \citep{Gamache2021}, and in \cite{Carlson1997} for NaCN. Our upper limits and relevant line parameters are given in Table~\ref{tab:limits}. These 3$\sigma$ upper limits on column density imply vapor masses of $<$11.9 kg, $<$24.5 kg, $<$66.7 kg, $<$37.1 kg, $<$2.9 kg, and $<$47.3 kg for SiO, SiS, SO, AlO, KCl, and NaCN, respectively, along the line of sight.

\begin{deluxetable*}{cccccccc}[h]
\tablenum{5}
\tablecaption{Upper Limits for Column Density of Gas-Phase Ejecta\label{tab:limits}}
\tablewidth{0pt}
\tablehead{
\colhead{Transition} & \colhead{$\nu$} & \colhead{$E_u/k_B$} & \colhead{$A_{ul}$} & \colhead{$\int T_b\,dv$} & \colhead{$<N_u>$} & \colhead{$<N>$} & \colhead{$N_{mol}$} \\
\colhead{} & \colhead{(GHz)} & \colhead{(K)} & \colhead{(s$^{-1}$)} & \colhead{(K \kms{})} & \colhead{(molecule cm$^{-2}$)} & \colhead{(molecule cm$^{-2}$)} & \colhead{(molecule)}
}
\startdata
SiO (\Ju{}=8--7) & 347.331 & 75.02 & $2.204\times10^{-3}$ & $<$ $2.1\times10^{-2}$ & $<$ $2.2\times10^9$ & $<$ $1.6\times10^{11}$ & $<$ $1.2\times10^{26}$  \\ 
SiS (\Ju{}=19--18) & 344.779 & 165.5 & $6.996\times10^{-4}$ & $<$ $1.5\times10^{-2}$  & $<$ $4.8\times10^9$  & $<$ $3.3\times10^{11}$  & $<$ $2.5\times10^{26}$ \\
SO ($J_K$=$8_8$--$7_7$) & 344.310 & 78.8 & $5.382\times10^{-4}$ & $<$ $1.3\times10^{-2}$  & $<$ $5.6\times10^9$  & $<$ $1.1\times10^{12}$  & $<$ $8.3\times10^{26}$ \\
AlO ($J_F$=$9_{12}$--$8_{11}$) & 344.476 & 82.8 & 4.768$\times10^{-3}$ & $<$ $1.7\times10^{-2}$ & $<$ $8.1\times10^8$ & $<$ $6.2\times10^{11}$ & $<$ $4.6\times10^{26}$ \\ 
KCl ($J$=45--44) & 344.820 & 321.3 & $2.504\times10^{-2}$ & $<$ $1.3\times10^{-2}$ & $<$ $1.2\times10^8$ & $<$ $3.2\times10^{10}$ & $<$ $2.3\times10^{25}$ \\
NaCN ($J_{Ka,Kc}$=$22_{3,20}$--$21_{3,19}$) & 344.269 & 211.7 &$1.784\times10^{-2}$ & $<$ $1.8\times10^{-2}$ & $<$ $2.4\times10^8$ & $<$ $7.9\times10^{11}$ & $<$ $5.8\times10^{26}$ \\
\enddata
\tablecomments{Frequencies, statistical weights, upper state energies, and Einstein-A coefficients retrieved from the LAMDA database \citep{Schoier2005} for SiO, SiS, and SO and from the CDMS database \citep{Endres2016} for all other species. $<N_u>$, $<N>$, and $N_{mol}$ are the column density in the upper transition state, total column density, and number of molecules, respectively, given for a $4\farcs68$ diameter aperture centered at the position of the asteroid system. All upper limits are 3$\sigma$.}
\end{deluxetable*}

\section{Conclusion}\label{sec:conclusion}
We conducted ALMA observations of the Didymos-Dimorphos system pre- and post-impact in support of the DART mission. Our pre-impact measurements provided the first measure of Didymos-Dimorphos' spectral emissivity at millimeter wavelengths, whose value was consistent with the handful of other siliceous and carbonaceous asteroids measured to date. Our post-impact measurements sampled thermal emission from Didymos-Dimorphos and from the dust ejecta liberated by the impact. By scaling our pre-impact measurements of Didymos-Dimorphos to the impact date, we isolated the emission owing to the ejecta and provided estimates of the ejecta dust mass throughout our observations between T$+$3.52 and T$+$8.60 hours post-impact. Our average ejecta masses of 1.3--$6.4\times10^7$ kg (depending on the material assumed) are consistent with pre-impact simulations, highlighting the effectiveness of the kinetic impactor method and the success of the first planetary defense test mission. As the ejecta were dominated by large particles, our ALMA measurements at $\lambda$ = 0.87 mm are the most sensitive to the total ejecta mass, and provide context and constraints for interpreting the properties of the ejecta as inferred from measurements at other wavelengths. Determining the mass and properties of the DART ejecta is critical for refining the kinetic impactor planetary defense technique for future applications, and our results highlight the sensitivity and capabilities of ALMA for supporting observations of spaceflight missions.

\begin{acknowledgments}
 This work was supported by the DART mission, NASA Contract No. 80MSFC20D0004. This work was supported by the Planetary Science Division Internal Scientist Funding Program through the Fundamental Laboratory Research (FLaRe) work package (NXR, SNM, MAC). SNM and CAT acknowledge support by NASA Planetary Science Division Funding through the Goddard Center for Astrobiology. JMTR acknowledges financial support from project PID2021-128062NB-I00 funded by Spanish MCIN/AEI/10.13039/501100011033. It makes use of the following ALMA data: ADS/JAO.ALMA \#2021.A.00013.S. ALMA is a partnership of ESO (representing its member states), NSF (USA), and NINS (Japan), together with NRC (Canada), MOST and ASIAA (Taiwan), and KASI (Republic of Korea) in cooperation with the Republic of Chile. The Joint ALMA Observatory is operated by ESO, AUI/NRAO, and NAOJ. The National Radio Astronomy Observatory is a facility of the National Science Foundation operated under cooperative agreement by Associated Universities, Inc. We thank two anonymous referees for their feedback, which we believe improved the manuscript.
\end{acknowledgments}

\appendix
\section{Self Calibration} \label{sec:selfcal}
Self-calibration was performed on both the pre- and post-impact observations to correct for residual phase and amplitude variations as allowed by the signal-to-noise ratio (S/N) of the data. Self-calibration was performed using CASA 6.5.2 and an automated self-calibration procedure\footnote{https://github.com/jjtobin/auto\_selfcal}. The automated self-calibration procedure utilizes best practices \citep[e.g.,][]{Brogan2018} and the auto-masking heuristics built into tclean \citep{Kepley2020}. Following the heuristics of the ALMA imaging pipeline, different auto-masking parameters were utilized for ALMA 12 m data vs.\ ACA data due to higher PSF sidelobes in the ACA data. The automated routine determined the optimal solution intervals to attempt for phase-only self-calibration given the setup of each observation. The data were more deeply cleaned with each successive solution interval as the data S/N improved, and the resulting model created for self-calibration by tclean can include fainter features without the risk of introducing artifacts into the model.

The first solution interval performed was a single solution over an entire execution block for each polarization independently, using gaincal parameters gaintype=`G' and combine=`scan', and solint=`inf'; this solution interval is referred to as `inf\_EB' (Table~\ref{tab:selfcal}). The routine first attempted to create these solutions on a per spectral window (spw) basis, but if there were too many flagged solutions, it fell back and used combine=`scan,spw', which was the case for all the observations presented here. This solution interval over the entire dataset corrected systematic errors in the data that could result from slight antenna position deviations and per-polarization phase offsets that might not have been completely taken out during standard calibration. The calibration table from the inf\_EB solution interval was pre-applied to all subsequent self-calibration solution intervals. This first solution interval was successful for all data presented here as evaluated by an increased S/N in the resulting images following self-calibration.

The next solution interval attempted was a single solution averaged over each scan in an observation, correcting for phase variations that occured between observations of the phase calibrator; this solution interval is referred to as `inf'. This solution interval varied in its actual length of time depending on the parameters of the observations and utilized the gaincal parameters combine=`spw', gaintype=`T', and solint=`inf' (Table~\ref{tab:selfcal}). Shorter solution intervals were attempted if the per-scan solutions were successful, and those solution intervals attempted to divide the scan as evenly as possible to include the same amount of data in each solution insofar as possible. Only one observation was able to use solution intervals shorter than a scan length.

Following the phase-only self-calibration, amplitude self-calibration was attempted if the per-scan phase-only solution interval was successful; this solution interval is referred to as `inf\_ap' (Table~\ref{tab:selfcal}). The amplitude self-calibration was run with both gaintables from the inf\_EB and  the shortest successful solution interval pre-applied, and then gaincal was executed with the parameters combine=`spw', gaintype=`T', solmode=`ap', solint=`inf', and solnorm=True. The use of solnorm=True means that the amplitude gains were normalized to 1 such that large changes in the overall flux density of the target were avoided.

The self-calibration gain tables were all applied using the CASA task applycal and we used the applymode=`calonly' parameter. This option prevented applycal from flagging data associated with flagged calibration solutions. The data associated with flagged calibration solutions were ``passed-through'' and did not have any self-calibration applied. This is because whenever gaincal flagged a solution for being less than the minimum S/N, it assigned that solution with the value 1+0$j$, which means that the amplitude and phase corrections were set to 1.0 and 0 degrees, respectively. A summary of the self-calibration application to the various executions is provided in Table~\ref{tab:selfcal}, where we identify the shortest successful solution interval for phase-only self-calibration and whether or not amplitude self-calibration was also applied. 

\begin{deluxetable}{ccccccc}[h]
\tablenum{A1}
\tablecaption{Self Calibration Summary Table}\label{tab:selfcal}
\tablewidth{0pt}
\tablehead{
      \colhead{Observation} & \colhead{UT Time} & \colhead{Best Successful} & \colhead{Pre-S.C. RMS\sups{(b)}} & \colhead{Pre-S.C.} & \colhead{Post-S.C. RMS\sups{(d)} }& \colhead{Post-S.C.}\\
      \colhead{} & \colhead{} & \colhead{Solution\sups{(a)}} & \colhead{(mJy beam$^{-1}$)} & \colhead{Dynamic Range\sups{(c)}} & \colhead{(mJy beam$^{-1}$)} & \colhead{Dynamic Range\sups{(e)}}\\
}
\startdata
      \hline
      \multicolumn{7}{c}{Pre-Impact Observations} \\
      1 & 03:49--03:55 & 100.8s, inf\_ap & 0.058 & 27 & 0.061 & 29\\
      \hline
      \multicolumn{7}{c}{Post-Impact 12 m Observations} \\
      1 & 02:18--03:15 & inf & 0.077 & 36 & 0.067 & 62\\
      2 & 03:34--04:31 & inf, inf\_ap & 0.061 & 57 & 0.049 & 95\\
      3 & 04:50--05:47 & inf & 0.054 & 83 & 0.048 & 101\\
      4 & 06:05--07:02 & inf, inf\_ap & 0.045 & 91 & 0.040 & 113\\
      \hline
      \multicolumn{7}{c}{Post-Impact ACA Observations} \\
      1 & 02:33--03:36 & inf\_EB & 0.768 & 5 & 0.706 & 7\\
      2 & 04:09--05:12 & inf & 0.468 & 10 & 0.447 & 14\\
      3 & 05:44--06:48 & inf & 0.405 & 12 & 0.379 & 15\\
      4 & 07:19--08:22 & inf\_EB & 0.432 & 11 & 0.424 & 12\\
\enddata
\tablecomments{\sups{(a)}Best successful solution interval given for phase self-calibration amd amplitude self-calibration. A single solution interval indicates that only phase self-calibration was applied (\eg{} all post-impact ACA observations), whereas a second solution interval indicates that amplitude self-calibration was also applied. \sups{(b)} Image RMS before self-calibration was applied. \sups{(c)} Image dynamic range (peak signal / RMS) before self-calibration was applied. \sups{(d)} Image RMS after self-calibration was applied. \sups{(e)} Image dynamic range after self-calibration was applied.
}
\end{deluxetable}


\pagebreak
\bibliography{DART}{}
\bibliographystyle{aasjournal}



\end{document}